%% file: muse_aspecs_stack_arXiv.tex
\documentclass[twocolumn]{aastex62}

\usepackage{overpic}
\usepackage{multirow}
\usepackage{xspace}

\newcommand{\cotwoone}{CO(2-1)\xspace}   % CO(2-1)
\newcommand{\cothreetwo}{CO(3-2)\xspace} % CO(3-2)

\newcommand{\subcubevel}{3000\xspace}    % The vel size of sub-cube (km/s)
\newcommand{\subcubexy}{11\xspace}       % The x & y size of sub-cube (arcsec)
\newcommand{\TotNCOTwoOne}{111\xspace}   % CO2-1: N of the total stacks
\newcommand{\DetNCOTwoOne}{10\xspace}          % CO2-1: N of the blind source detection
\newcommand{\TotEXNCOTwoOne}{101\xspace} % CO2-1: N of the stacks
\input{CO2-1_table_values}

\input{CO3-2_table_values}

\input{CO4-3_table_values}

\input{CO5-4_table_values}

\input{CO6-5_table_values}

\shorttitle{ASPECS CO stacking analysis}
\shortauthors{Inami et al.}
\begin{document}
       
\title{The ALMA Spectroscopic Survey in the HUDF: Constraining the
  Molecular Content at $\log{(M_*/M_\odot)} \sim 9.5$ with CO stacking
  of MUSE detected $z\sim1.5$ Galaxies}

\correspondingauthor{Hanae inami}
\email{hanae@hiroshima-u.ac.jp}

\author[0000-0003-4268-0393]{Hanae Inami} 
\affiliation{Hiroshima Astrophysical Science Center, 
             Hiroshima University, 
             1-3-1 Kagamiyama, 
             Higashi-Hiroshima, 
             Hiroshima 739-8526, Japan}

\author[0000-0002-2662-8803]{Roberto Decarli}
\affiliation{INAF–Osservatorio di Astrofisica e Scienza dello Spazio,
  via Gobetti 93/3, I-40129, Bologna, Italy}

\author[0000-0003-4793-7880]{Fabian Walter}
\affiliation{Max-Planck-Institut f\"ur Astronomie, K\"onigstuhl 17,
             D-69117 Heidelberg, Germany}
\affiliation{National Radio Astronomy Observatory, Pete V. Domenici
  Array Science Center, P.O. Box O, Socorro, NM 87801, USA}

\author[0000-0003-4678-3939]{Axel Weiss}
\affiliation{Max-Planck-Institut für Radioastronomie, Auf dem Hügel
  69, D-53121 Bonn, Germany}

\author[0000-0001-6647-3861]{Chris Carilli}
\affiliation{National Radio Astronomy Observatory, Pete V. Domenici
  Array Science Center, P.O. Box O, Socorro, NM 87801, USA}
\affiliation{Battcock Centre for Experimental Astrophysics, Cavendish
  Laboratory, Cambridge CB3 0HE, UK}

\author[0000-0002-6290-3198]{Manuel Aravena}
\affiliation{N\'ucleo de Astronom\'ia de la
  Facultad de Ingenier\'ia y Ciencias, Universidad Diego Portales,
  Av. Ej\'ercito Libertador 441, Santiago, Chile}

\author[0000-0002-3952-8588]{Leindert Boogaard}
\affiliation{Leiden Observatory, Leiden
  University, P.O. Box 9513, NL-2300 RA Leiden, The Netherlands}

\author[0000-0003- 3926-1411]{Jorge Gonz\'alez-L\'opez}
\affiliation{N\'ucleo de Astronom\'ia de la
  Facultad de Ingenier\'ia y Ciencias, Universidad Diego Portales,
  Av. Ej\'ercito Libertador 441, Santiago, Chile}
\affiliation{Instituto de Astrof\'isica, Facultad de F\'isica,
  Pontificia Universidad Cat\'olica de Chile Av. Vicu\~na Mackenna 4860,
  782-0436 Macul, Santiago, Chile}

\author[0000-0003-1151-4659]{Gerg\"o Popping}
\affiliation{European Southern Observatory, Karl-Schwarzsschild-Str. 2, D-85748, Garching}

\author[0000-0001-9759-4797]{Elisabete da Cunha}
\affiliation{International Centre for Radio Astronomy Research,
  University of Western Australia, 35 Stirling Hwy, Crawley, WA 6009, Australia}
\affiliation{Research School of Astronomy and Astrophysics,
  The Australian National University, Canberra, ACT 2611, Australia}
\affiliation{ARC Centre of Excellence for All Sky Astrophysics in 3 Dimensions (ASTRO 3D)}

\author{Roland Bacon}
\affiliation{Univ. Lyon 1, ENS de Lyon, CNRS, Centre de Recherche
  Astrophysique de Lyon (CRAL) UMR5574, F-69230 Saint-Genis-Laval,
  France}

\author[0000-0002-8686-8737]{Franz Bauer}
\affiliation{Instituto de Astrof\'isica, Facultad de F\'isica,
  Pontificia Universidad Cat\'olica de Chile Av. Vicu\~na Mackenna 4860,
  782-0436 Macul, Santiago, Chile}
\affiliation{National Radio Astronomy Observatory, Pete V. Domenici
  Array Science Center, P.O. Box O, Socorro, NM 87801, USA}
\affiliation{Space Science Institute, 4750 Walnut Street, Suite 205,
  Boulder, CO 80301, USA}

\author[0000-0003-0275-938X]{Thierry Contini}
\affiliation{Institut de Recherche en Astrophysique et Plan\'etologie (IRAP),
  CNRS, 14, avenue Edouard Belin, F-31400 Toulouse, France}
\affiliation{Universit\'e de Toulouse, UPS-OMP, Toulouse, France}

\author[0000-0002-3583-780X]{Paulo C. Cortes}
\affiliation{Joint ALMA Observatory - ESO, Av. Alonso de C\'ordova, 3104, Santiago, Chile}
\affiliation{National Radio Astronomy Observatory, 520 Edgemont Rd, Charlottesville, VA, 22903, USA}
  
\author[0000-0003-2027-8221]{Pierre Cox}
\affiliation{Institut d'astrophysique de Paris, Sorbonne
  Universit\'e, CNRS, UMR 7095, 98 bis bd Arago, F-7014 Paris, France}

\author[0000-0002-3331-9590]{Emanuele Daddi}
\affiliation{Laboratoire AIM, CEA/DSM-CNRS-Universite Paris Diderot,
  Irfu/Service d'Astrophysique, CEA Saclay, Orme des Merisiers,
  F-91191 Gif-sur-Yvette cedex, France}

\author[0000-0003-0699-6083]{Tanio D\'iaz-Santos}
\affiliation{N\'ucleo de Astronom\'ia de la
  Facultad de Ingenier\'ia y Ciencias, Universidad Diego Portales,
  Av. Ej\'ercito Libertador 441, Santiago, Chile}
\affiliation{Chinese Academy of Sciences South America Center for
  Astronomy (CASSACA), National Astronomical Observatories, CAS,
  Beijing 100101, China}
\affiliation{Institute of Astrophysics, Foundation for Research and
  Technology—Hellas (FORTH), Heraklion, GR-70013, Greece}

\author{Melanie Kaasinen}
\affiliation{Max-Planck-Institut f\"ur Astronomie, K\"onigstuhl 17,
  D-69117 Heidelberg, Germany}
\affiliation{Universit\"{a}t Heidelberg, Zentrum f\"{u}r
  Astronomie, Institut f\"{u}r Theoretische Astrophysik,
  Albert-Ueberle-Stra\ss e 2, D-69120 Heidelberg, Germany}

\author[0000-0001-9585-1462]{Dominik A. Riechers}
\affiliation{Department of Astronomy, Cornell University,
             Space Sciences Building, Ithaca, NY 14853, USA}
\affiliation{Max-Planck-Institut f\"ur Astronomie, K\"onigstuhl 17,
             D-69117 Heidelberg, Germany}

\author{Jeff Wagg}
\affiliation{SKA Organization, Lower Withington Macclesfield, Cheshire
  SK11 9DL, UK}

\author[0000-0001-5434-5942]{Paul van der Werf}
\affiliation{Leiden Observatory, Leiden University, P.O. Box 9513,
  NL-2300 RA Leiden, The Netherlands}

\author{Lutz Wisotzki}
\affiliation{Leibniz-Institut f\"ur Astrophysik Potsdam (AIP), An der
  Sternwarte 16, 14482 Potsdam, Germany}

% \author{August Muench}
% \affiliation{American Astronomical Society \\
% 2000 Florida Ave., NW, Suite 300 \\
% Washington, DC 20009-1231, USA}
% \collaboration{(AAS Journals Data Scientists collaboration)}

% \author{Butler Burton}
% \affiliation{National Radio Astronomy Observatory}
% \affiliation{AAS Journals Associate Editor-in-Chief}
% \nocollaboration

% \author{Amy Hendrickson}
% \altaffiliation{Creator of AASTeX v6.2}
% \affiliation{TeXnology Inc.}
% \collaboration{(LaTeX collaboration)}

% \author{Julie Steffen}
% \affiliation{AAS Director of Publishing}
% \affiliation{American Astronomical Society \\
% 2000 Florida Ave., NW, Suite 300 \\
% Washington, DC 20009-1231, USA}

% \author{Jeff Lewandowski}
% \affiliation{IOP Senior Publisher for the AAS Journals}
% \affiliation{IOP Publishing, Washington, DC 20005}



%% Up tp 250 words
\begin{abstract}
  We report molecular gas mass estimates obtained from a
  stacking analysis of CO line emission in the ALMA Spectroscopic
  Survey (ASPECS) using the spectroscopic redshifts from
  the optical integral field spectroscopic survey by the Multi Unit
  Spectroscopic Explorer (MUSE) of the {\it Hubble} Ultra Deep Field
  (HUDF).  Stacking was performed on subsets of the sample of galaxies
  classified by their stellar mass and position relative to the
  main-sequence relation (on, above, below).  Among all the CO
  emission lines, from \cotwoone to CO(6-5), with redshifts accessible
  via the ASPECS Band~3 and the MUSE data, \cotwoone provides the
  strongest constraints on the molecular gas content.  We detect
  \cotwoone emission in galaxies down to stellar masses of
  $\log{(M_*/M_\odot)}=10.0$.  Below this stellar mass, we present a
  new constraint on the molecular gas content of $z\sim1.5$
  main-sequence galaxies by stacking based on the MUSE detections.  We
  find that the molecular gas mass of main-sequence galaxies
  continuously decreases with stellar mass down to
  $\log{(M_*/M_\odot)}\approx9.0$.  Assuming a metallicity-based
  CO--to--$\rm H_2$ conversion factor, the molecular
    gas-to-stellar mass ratio from $\log{(M_*/M_\odot)}\sim9.0$ to
    $\sim10.0$ does not seem to decrease as fast as for
    $\log{(M_*/M_\odot)}>10.0$, which is in line with simulations and
  studies at lower redshift.
  The inferred molecular gas density
  $\rho{\rm (H_2)}=(0.49\pm0.09)\times10^8\,{\rm
    M_\odot\,Mpc^{-3}}$ of MUSE-selected galaxies at $z\sim1.5$ is
  comparable with the one derived in the HUDF with a different CO
  selection.  Using the MUSE data we recover most of the CO emission
  in our deep ALMA observations through stacking, demonstrating the
  synergy between volumetric surveys obtained at different wavebands.
\end{abstract}



\section{Introduction} \label{sec:intro}

Stars form inside dense molecular gas clouds. It is thus critical to
reveal how much molecular gas exists in galaxies to characterize
galaxy formation and evolution
\citep[e.g.,][]{Kenn12,Cari13b,Tacc20,Hodg20}.  It is well established
that the cosmic star formation rate density increased from the early
stages of the Universe towards its peak around $z \sim 1-3$, after
which it progressively decreased until the current epoch
\citep{Mada14}.  A broad consensus is emerging on the cause of this
growth, peak, and decline of the star formation history over cosmic
time via measurements of the gas that fuels star formation
\citep[e.g.,][]{Walt14,Deca16a,Scov17b,Deca19,Riec19,Liu19,Magn20,Lenk20,Tacc20}. This
evolution could be due to either the available supply of molecular gas
for forming stars, a mechanism that causes high efficiency in star
formation such as galaxy mergers, or a combination of these processes.

Most of the star formation in the Universe occurs in galaxies residing
on the so-called ``main sequence'', a tight correlation between the
star formation rate (SFR) and stellar mass ($M_*$) of star-forming
galaxies
\citep[e.g.,][]{Noes07,Salm12,Whit12,Whit14,Schr15,Pope19a,
  Pope19b}. This correlation is observed at redshifts up to at least
$z\sim6.5$ \citep[e.g.,][]{Spea14,Salm15}. In other words, most star formation in
the Universe is long-lasting and evolves steadily, supporting the idea
that it is predominantly regulated by the gas accretion of the
available fuel supply and feedback processes, rather than stochastic
events like galaxy mergers \citep{Lill13,Deke13,Tacch16}.  Such
stochastic events cause enhanced star formation activity, which
elevates SFRs significantly above the main-sequence relation. However,
these galaxies, referred to as starbursts are in the minority
\citep[e.g.,][]{Rodi11,Sarg12,Lutz14}.  To understand these two star
formation modes and the efficiency of the star formation process, it
is essential to determine the $\rm H_2$ gas supply and deficiency. The
most common tracer of the molecular gas, the fuel of star formation,
is line emission from carbon monoxide ($^{12}{\rm CO}$) rotational
transitions at low excitation \citep{Bola13b,Cari13b}.
  
The connection between molecular gas content, stellar mass, and SFR
for $z > 1$ galaxies on and above the main-sequence relation has been
investigated in various targeted studies
\citep[e.g.,][]{Magd12b,Sant14,Scov16,Tacc13,Tacc18}. However, the
targets have so far been limited to massive ($> 10^{10} \, M_\odot$)
and highly star-forming ($\gtrsim 50 M_\odot \, {\rm yr^{-1}}$)
galaxies.  Little is known about the molecular gas reservoirs in
galaxies either below the main sequence or at modest stellar masses,
mostly because of sensitivity limits. 

The ALMA Spectroscopic Survey in the {\it Hubble} Ultra-Deep Field
\citep[ASPECS,][]{Walt16} has been conducted as a spectroscopic survey
over the entire frequency range of ALMA Bands~3 and 6 in the {\it Hubble}
Ultra Deep Field \citep[HUDF;][]{Beck06} to perform an unbiased search
for multiple rotational transitions of CO emission
\citep{Walt16,Gonz19}. Spectral line scans have an advantage in
assessing the molecular gas content based on a complete line
flux-limited sample without any target preselections. \cite{Gonz19}
conducted a blind search of line and continuum sources directly in the
ASPECS data cube (Band~3) and evaluated its completeness.  Using these
reliable CO detections, \cite{Deca19} constructed CO luminosity
functions and presented the evolution of the cosmic gas mass density.
A census of the molecular gas content of galaxies that have direct CO
detections is shown and discussed in \cite{Arav19}.  These gas
measurements were compared with model predictions from cosmological
simulations and semi-analytical models in \cite{Popp19}.
\cite{Uzgi19} performed a power spectrum analysis and probed CO
emission at $1 \lesssim z \lesssim 4$ below the sensitivity limit of
individual detections and gave a constraint on missing CO emission
from individually undetected galaxies.

Here, we maximize the sensitivity of the ASPECS data to detect CO
emission by stacking ALMA spectra using the Band~3 data. Our stacking
analysis is based on optical spectroscopic redshifts from another
large, unbiased, blind spectroscopic survey in the HUDF carried out by
the integral field unit (IFU) instrument MUSE (Multi Unit
Spectroscopic Explorer) on the Very Large Telescope
\citep[][]{Baco17}. The combination of the three-dimensional (3D) data
obtained by both the ASPECS and MUSE surveys not only made the
stacking analysis possible, but also enabled a direct comparison
between the molecular gas properties and rest-frame
optical/ultraviolet properties \citep{Boog19}.

This paper is structured as follows: we first introduce the
observations and data taken with the ASPECS and MUSE HUDF surveys and
the ancillary data in \S\ref{sec:sample}. We describe the method of
the stacking analysis along with the sample used for stacking in
\S\ref{sec:method}.  In \S\ref{sec:results}, the stacked CO spectra
are presented. We then convert the measured CO emission to molecular
gas mass and discuss the gas mass content in \S\ref{sec:discussion}.
We summarize and conclude our findings in \S\ref{sec:summary}.  A flat
$\Lambda$CDM cosmology with $H_0 = 70 \, {\rm km\,s^{-1}\,Mpc^{-1}}$,
$\Omega_\Lambda = 0.7$, and $\Omega_m = 0.3$ is adopted throughout
this paper.

\section{The ASPECS and MUSE Datasets}\label{sec:sample}

The MUSE survey covered the entire area of the HUDF \citep{Baco17},
whereas the ASPECS LP observed almost the entire region of the {\it Hubble}
eXtremely Deep Field \citep[XDF;][]{Illi13}.  MUSE is an optical IFU
\citep{Baco15} on the Very Large Telescope (VLT) Yepun (UT4) of the
European Southern Observatory (ESO) with a wide field-of-view (FoV,
$1\arcmin \times 1\arcmin$), high sensitivity, wide wavelength
coverage ($4650-9300\,{\rm \AA}$), and high spectral resolution
($R\sim3000$). In particular, the large size of its FoV facilitates
spectroscopic redshift surveys without requiring any target
preselection, achieving a spatially homogeneous spectroscopic
completeness. It offers redshift determination at $z=0-6.5$ based on
rest-frame ultraviolet and optical emission and absorption lines.

\begin{figure*}
  \begin{center}
    \includegraphics[angle=0,width=1.0\textwidth]{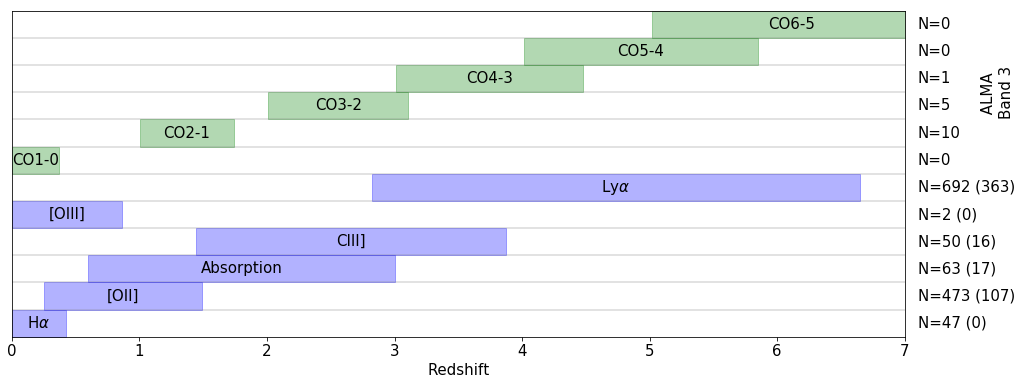}
    \caption{The redshift coverages of the MUSE (blue) and ASPECS
      (green) HUDF surveys. The numbers of ASPECS directly detected
      sources and MUSE sources with a secure redshift measurement are
      listed on the right hand side \citep[There are also nine
        and two sources classified as ``stars'' and ``others'',
        respectively,][]{Inam17}. The numbers in parentheses are the
      MUSE sources whose locations are in the region of LP primary
      beam response $>50\%$ for each CO emission. The absorption
        features detected with MUSE include \ion{C}{4},
        \ion{Fe}{2}, \ion{Mg}{2}. The analysis presented in this work
      focuses on the stacking frequencies that correspond to the
      \cotwoone, \cothreetwo, CO(4-3), CO(5-4), and CO(6-5) lines. }
    \label{fig:MUSE_ASPECS}
  \end{center}
\end{figure*}

Complementing the MUSE spectroscopic survey, the ASPECS line scan
survey in ALMA Band~3 (3mm) can detect multiple CO rotational
transition lines from $J=1-0$ to $J=6-5$ in the redshift range of
$z=0-7$ \citep[with gaps\,\footnote{The ASPECS Band~6 (1mm) survey
  covers most of these gaps by observing higher-$J$ CO lines.} at
$0.37 < z < 1.00$ and $1.74 < z < 2.00$,][]{Walt16}.  The MUSE
redshift coverage overlaps well with the redshift coverage of the
ASPECS Band~3 line scan survey at $z > 1$
(Figure~\ref{fig:MUSE_ASPECS}), except for a small gap at
$z=1.74-2.00$.

% Fig.1

In this paper, we concentrate on the ASPECS Band~3 data where the full
MUSE redshift coverage can be exploited (see
\S\ref{subsec:subsamples}).  The significant redshift overlap of these
two unbiased spectroscopic surveys in the same field is beneficial for
performing stacking analyses on ASPECS CO spectral lines based on the
optically determined MUSE redshifts.  In this section, we will briefly
present the survey designs of the HUDF conducted by ASPECS and MUSE.

\subsection{ASPECS observations and data}

The detailed survey strategy and data reduction of the ASPECS Pilot
and Large Program surveys are presented in \cite{Walt16} and
\cite{Gonz19}, respectively. The observational setups of the Large
Program (LP) survey were the same as the Pilot survey, except its
coverage is extended to $\sim 5 \, {\rm arcmin}^2$.  In this work, we
used the combined data of the ASPECS Pilot and LP surveys taken with
ALMA Band~3. ASPECS LP carried out a full frequency scan in Band~3
($84-115$\,GHz) over the XDF, which resides in the HUDF. With a total
of 17 pointings centered at (RA, Dec) $=$ (03:32:38.5, -27:47:00),
which completely cover the Pilot region, the total area with a primary
beam response $>50\%$ in the LP survey is $4.6\,{\rm arcmin}^2$ at
$\approx 99.5\,{\rm GHz}$ (the central frequency of Band~3).

The \textsc{CASA} (Common Astronomy Software Applications) software
was used to calibrate and image the data.  With the C40-3 array
configuration, we obtained a synthesized beam size of
$1.75\arcsec \times 1.49\arcsec$ with a position angle $91.5\deg$ at
$\approx 99.5\,{\rm GHz}$ by using natural weighting in
  \textsc{CASA} when imaging the data. The frequency channel was
rebinned to 7.813\,MHz ($23.5\,{\rm km\,s^{-1}}$ at 99.5\,GHz). The
sensitivity for this channel bin size was
$\sim 0.2\,{\rm mJy\,beam^{-1}}$ throughout the entire scanned
frequency range. The $5\sigma$ \cotwoone line sensitivity is
$> 1.4 \times 10^9 \, {\rm K\,km\,s^{-1}\,pc^2}$ assuming a line width
of $200\,{\rm km\,s^{-1}}$. Based on assumptions made in this work
(see \S\ref{subsec:gas_mass}), the corresponding molecular gas
($\rm H_2$) limit is $\ga 6.8 \times 10^9 \, M_\odot$.  For our main
target emission line, \cotwoone, the ALMA Band~3 scan offers redshift
coverage of $z=1.0059-1.7387$.

Using the same data cube, \cite{Gonz19} reported CO emission line
detections from an unbiased blind search without prior knowledge of
source positions and observed CO frequencies. They found 16 high
significance CO emitting sources, among which 11 were identified as
\cotwoone emission.  We refer the readers to \cite{Gonz19} for
comprehensive discussions on the detection methods.  \cite{Arav19}
assessed the molecular gas properties of these sources, which will be
used for comparison in this paper.

\subsection{MUSE observations and data}\label{subsec:muse}

The MUSE-HUDF deep survey was conducted as a two layer survey of
different depths \citep{Baco17}. The $3\arcmin \times 3\arcmin$ deep
survey observed the entire HUDF region, whereas the
$1\arcmin \times 1\arcmin$ ultra deep survey was carried out near the
center of the deep survey area. The $\approx 10$\,h and
$\approx 31$\,h exposure times, respectively, reached $3\sigma$
emission line flux limits of $3.1$ and
$1.5 \times 10^{-19} \, {\rm erg\,s^{-1}\,cm^{-2}}$ at
$\sim 7000\,{\rm \AA}$.

The MUSE spectra were extracted with two methods: prior extractions
based on the UVUDF catalog \citep{Rafe15} and a blind search for
emission lines in the data cubes. In the former case, the coordinates
for the source extraction were from UVUDF\,\footnote{In cases where
  MUSE could not spatially resolve the {\it HST}-detected sources,
  these sources were ``merged'' into a single MUSE object. Its new
  coordinates are the {\it HST} F775W flux-weighted center of all the
  merged objects.}, whereas for the latter, the coordinates were
determined from the MUSE data. For a more detailed description of the
survey and data treatments, see \cite{Baco17}. Following
\cite{Dunl17}, we corrected the known systematic offset of the {\it
  Hubble} Space Telescope ({\it HST}) positions to match the radio
astrometric reference frame by applying $+0.279\arcsec$ in Declination
(Dec) and $-0.076\arcsec$ in Right Ascension (R.A.).

The MUSE spectroscopic redshifts were measured from the spectral
features in the extracted spectra. Each redshift has an associated
confidence level of 3 (secure redshift, determined by multiple
spectral lines), 2 (secure redshift, determined by a single spectral
line), or 1 (possible redshift, determined by a single spectral line
whose spectral identification remains uncertain).  The typical MUSE
redshift uncertainty is $\sigma_z = 0.00012(1+z)$ or
$\sigma_v \approx 40 \, {\rm km \, s^{-1}}$, which is smaller than the
typical line width of CO emission.  We refer to \cite{Inam17} for
details about the redshift determination and redshift catalogs of the
MUSE-HUDF field. In this work, we only use the reliable MUSE redshifts
of confidence levels 2 and 3.

The MUSE-HUDF survey obtained 1338 reliable redshifts in total, a
factor of eight increase over the previously available spectroscopic
redshifts in this field.  The simultaneous wavelength coverage of
$4650-9300\,{\rm \AA}$ and the spectral resolution of $R\sim3000$ of
MUSE allowed detections and unambiguous identifications of major
rest-frame ultraviolet and optical emission lines, including
H$\alpha$, [\ion{O}{2}], [\ion{O}{3}], and Ly$\alpha$.  These lines
were used to determine redshifts over the range $0 < z <
6.5$. Although more difficult to probe, the redshift range
$z \sim 1.5-3$ can be covered by \ion{C}{3}] and absorption features
\citep[e.g., \ion{C}{4}, \ion{Fe}{2}, \ion{Mg}{2}; see Figure~13
of][]{Inam17}.

Among the spectroscopic redshifts assessed in the MUSE HUDF field,
503 sources with a confidence level of 2 or higher lie within
the ASPECS survey region (LP primary beam response $> 50\%$). Out of
these sources, 107 sources are [\ion{O}{2}] emitters covering
$z=0.25-1.5$ and 363 sources are Ly$\alpha$ emitters covering
$z=2.8-6.6$. For the analysis presented in the main part of this
paper, we took advantage of the prevalent [\ion{O}{2}] line detections
and their redshift overlap with the \cotwoone selection function in
ALMA Band~3 (Figure~\ref{fig:MUSE_ASPECS}) to perform a stacking
analysis. In the redshift range where \cotwoone can be detected, the
MUSE spectroscopic redshift sources with absorption features and
\ion{C}{3}] emission also contributed to the stacking, although the
number was small (Figure~\ref{fig:MUSE_ASPECS}). We also attempted to
stack spectra to detect higher-$J$ CO lines.  The MUSE redshifts used
in higher-$J$ CO stacked spectra were mostly measured using the
Ly$\alpha$ line, which is known to be offset from the systemic
redshift for a few hundred $\rm km\,s^{-1}$
\citep[e.g.,][]{Shap03}. We have applied a correction to this offset
(\S\ref{subsec:results_hiJ}).

\subsection{Ancillary data and physical parameters derived from SED fitting}\label{subsec:sed}

Owing to the same HUDF coverage of the MUSE and ASPECS observations,
there are abundant ancillary data available. We assembled optical and
near-infrared photometric data from \cite{Skel14}. These photometry
catalogs and data include ultraviolet to infrared from the {\it HST},
various ground-based telescopes, and all of the {\it Spitzer} IRAC
channels.

Based on this photometric dataset, we inferred physical parameters
such as star formation rate (SFR) and stellar mass ($M_*$) via
modeling with the high redshift extension of the Spectral Energy
Distribution (SED) fitting code \texttt{MAGPHYS} \citep{daCu08,
  daCu15}. In addition to {\it Spitzer}/MIPS and {\it
  Herschel}/PACS, the ALMA 1.2 and 3\,mm photometry
  from the ASPECS data \citep{Gonz19,Gonz20} were also used for the
SED fitting for a subset of the sources whose CO or continuum emission
was detected. The same procedure was carried out in the other ASPECS
work \citep[][]{Arav20,Boog20}.  The SED fits were computed
based on the MUSE spectroscopic redshift.  See \cite{Boog19} for 
  the detailed process of the SED fitting.

\section{Methods} \label{sec:method}

\subsection{ALMA spectral extraction and stacking}\label{subsec:extract}

From the ALMA data, we first extracted a sub-cube centered at
the position of each MUSE source with a secure spectroscopic redshift.
This sub-cube has a size of
$\subcubexy\arcsec \times \subcubexy\arcsec \times \subcubevel\,{\rm
  km\,s^{-1}}$. The primary beam correction using the combined Pilot
and LP primary beam response was applied after the sub-cube
extraction.

The uncertainty of the extracted spectra was calculated based on the
region in the data cube where the combined beam response is $> 99\%$
of the peak sensitivity at the phase center.  The uncertainty for each
spectrum is then scaled by the combined primary beam response of the
location of the objects under consideration.

To perform stacking, the channel frequencies of the extracted spectra
were converted to the rest-frame, then the spectra were resampled onto
a common frequency grid. In the rest-frame, a CO line detected at the
lower frequency end of the spectrum (i.e., galaxies at higher
redshift) has a coarser spectral sampling due to the $(1+z)$
correction of the redshift.  Thus, the common frequency grid was given
at the coarsest sampling, corresponding to the CO line detected at the
lowest frequency end of Band~3.  In the case of \cotwoone, we set the
final spectral range and the channel width of the extracted spectra to
be $\subcubevel \, {\rm km\,s^{-1}}$, centered on the \cotwoone
rest-frame frequency ($230.538\,$GHz), and $27.8 \, {\rm km\,s^{-1}}$
($21.4 \, {\rm MHz}$), respectively. This resulted in 108 channels in
the extracted spectra. The final extracted rest-frame frequency range
is $229.395 - 231.682\,$GHz ($\pm 1500 \, {\rm km\,s^{-1}}$). The same
procedure is adopted for stacking high-$J$ CO lines.

This conversion was implemented by first shifting the channel
frequencies of the spectra to the rest-frame, according to the MUSE
redshift.  We then applied a Gaussian decimation filter to the
rest-frame spectra via the Fourier plane.  This filter removes fine
scale (high sampling frequency) noise in the spectrum which would
otherwise be aliased into spurious noise when sub-sampled to the lower
target resolution \citep{Lyons11}.  A better signal-to-noise ratio
(SNR) was thus obtained when we binned all spectra onto the coarser
common frequency grid in the later step.  In the Fourier plane, the
width of the Gaussian filter was set to give an attenuation of 40\,dB
at the Nyquist folding frequency of half a cycle per channel of the
new channel width (0.5 cycles per $27.8 \, {\rm km\,s^{-1}}$).  This
value is a reasonable compromise between the desirable suppression of
aliased noise and the corresponding (minor) reduction in the
resolution of the sub-sampled spectra. The Gaussian kernel was scaled
to be unity at the origin of the Fourier plane to guarantee flux
conservation. In addition, the kernel was given an odd number of
elements, placed symmetrically around the origin, to prevent spectral
shifts during the convolution.  The effect of this filter was
equivalent to performing a linear convolution of the spectrum with an
area-normalized Gaussian of FWHM $63.2\,{\rm km\,s^{-1}}$.

Next, we interpolated the filtered spectrum onto the frequency grid of
the stacked spectra.  The mean of the resulting spectra was then
calculated to obtain the final stacked spectrum \,\footnote{The median
  stacking also produced similar results, but here we adopt the mean
  stacking for simpler noise estimates and for a better treatment when
  there are only two samples to stack.}. We did not apply any
weighting when performing the stack.
  
We also carried out two different random extractions following the
same procedure described above, to produce random stacked spectra for
comparisons.  One random extraction involved assigning random
redshifts to each spectrum before combining them at the location of
the known MUSE position. The other involved using the correct MUSE
redshifts, but extracting the spectra at random positions within the
region where the LP primary beam response is $> 50\%$, instead of at
the known source position.  The number of randomized extractions is
the same as the number of CO emission samples.  Neither of these
methods should produce stacked spectra with real features.  This
random spectral extraction highlights which features in the real
stacked spectra should be discounted as noise.

\begin{figure}
  \begin{center}
    \includegraphics[angle=0,width=0.5\textwidth]{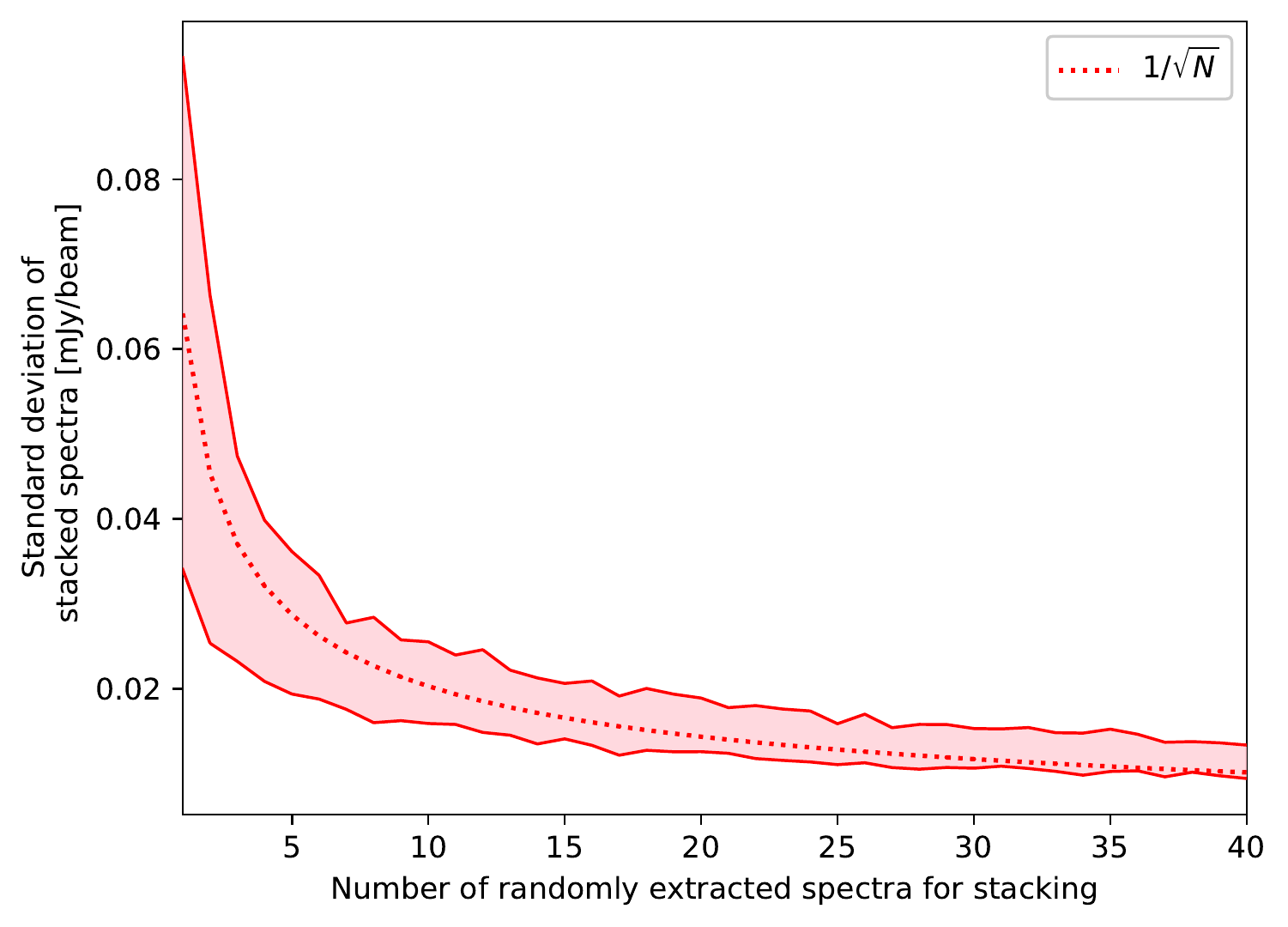}
    \caption{Reduction of standard deviation of stacked spectra
        (100 realizations) with increasing number of stacked spectra
        (red region). The stacked spectra used here were extracted
        randomly from the ASPECS data cube. The red dotted line
        represents $1/\sqrt{N}$, where $N$ is the number of the
        stacked spectra.}
    \label{fig:stack_rms}
  \end{center}
\end{figure}

In Figure~\ref{fig:stack_rms}, we show the standard deviation of
randomly stacked spectra against the number of the stacks. For this
plot, we randomize both redshift and position for the spectral
extractions.  The standard deviation of stacked spectra roughly
decreases with $1/\sqrt{N}$.

% Fig.2

\subsection{Sample selection}\label{subsec:subsamples}

We took advantage of the significant redshift overlap between the
ASPECS and MUSE surveys to carry out a stacking analysis on CO line
emission: we focused on the ASPECS Band~3 data where the full MUSE
redshift coverage can be exploited.  In particular, over the range
$1.0059 \leq z \leq 1.7387$, the ASPECS survey has the highest
sensitivity for detecting \cotwoone among all of the observable CO
emission features \citep[Figure~9 of][]{Boog19}. In addition, this
redshift range is where MUSE is efficient at detecting spectral
features (up to $z\sim1.5$ for [\ion{O}{2}]) as shown in
Figure~\ref{fig:MUSE_ASPECS}.

We first selected a subset of the MUSE sources for stacking \cotwoone
emission in the ASPECS Band~3 data cube. The criteria were the
following: (1) $1.0059 \leq {\rm MUSE \ spec-}z \leq 1.7387$, (2) MUSE
spec$-z$ confidence levels $\geq 2$, and (3) location within ALMA
Band~3 LP primary beam response $\geq 50\%$. For high-$J$ CO emission,
we used the same selection criteria, except for the redshift
range. The ranges were $z=2.0088-3.1080$ for \cothreetwo,
$z=3.0115-4.4771$ for CO(4-3) $z=4.0142-5.8460$ for CO(5-4)
$z=5.0166-7.2146$ for CO(6-5).

The resulting number of MUSE sources for stacking \cotwoone was
\TotNCOTwoOne in total~\footnote{There are three galaxies (MUSE
  IDs 6314, 6450, and 6530) which meet the selection
    criteria, but their stellar masses and SFRs cannot be constrained
  because they have no optical counterpart on which to
  perform an SED fit. These galaxies are found by MUSE (no {\it HST}
  counterpart in the UVUDF catalog due to blending). These objects are
  not included in our sample of \TotNCOTwoOne.}.  For 104
sources, the MUSE spectroscopic redshift identifications were based on
[\ion{O}{2}], five sources used absorption features, one source used
\ion{C}{3}], and one quasar had strong \ion{Mg}{2} emission.
% TYPE 3 = MUSE ID 83 (FeII & MgII), 886 (MgII & Ca), 972, 996, 1001
% TYPE 4 = MUSE ID 42 (CIII])
% TYPE 7 = MUSE ID 872 (quasar, strong MgII emission) 
Among these \TotNCOTwoOne sources, \DetNCOTwoOne
sources~\footnote{Four of them detected by the CO blind search,
  ASPECS-LP-3mm.02, 04, 05, and 08, have a MUSE redshift confidence
  level of 1, but they are included in this work.}  were identified
with \cotwoone emission by the ASPECS blind search: MUSE IDs 8
(ASPECS-LP-3mm.06), 16 (3mm.11), 924 (3mm.14), 925 (3mm.16), 996
(3mm.02), 1001 (3mm.05), 1011 (3mm.10),
1117 (3mm.04), 6415 (3mm.08), and 6870 (3mm.15) \citep{Boog19,Arav19}.

\begin{figure}
  \begin{center}
    \includegraphics[angle=0,width=0.45\textwidth]{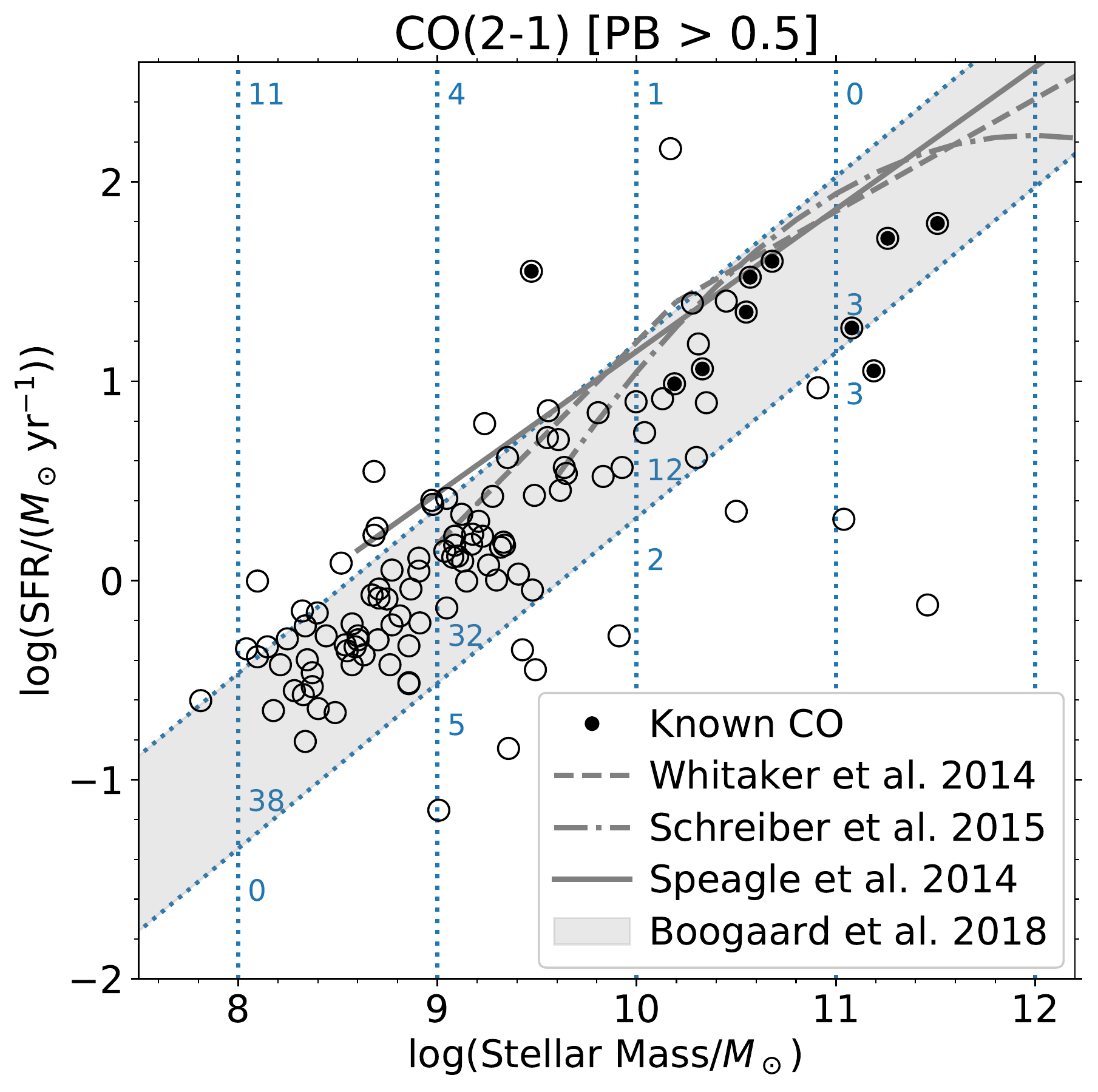}
    \caption{ SFR-$M_*$ relation of our \cotwoone stacking sample.
      The gray filled band shows the main-sequence relation and its
      intrinsic scatter for MUSE-detected galaxies \citep[at
      $z=1.43$,][]{Boog18}. The gray solid, dashed and dashed dotted
      lines are the main-sequence relations from \cite{Spea14}
      \citep[which is used for PHIBSS2][see also
      Appendix~\ref{app:Spea14}]{Tacc18}, \cite{Whit14} and
      \cite{Schr15}, respectively.  The blue vertical and diagonal
      dotted lines show the divisions of the bins that we used for
      stacking. The numbers at the top left of each grid are the
      counts of galaxies in the bin.  The filled circles indicate the
      galaxies which have \cotwoone line detections in the ASPECS
      blind search (see \S\ref{subsec:subsamples}).
    }
    \label{fig:SFR-M}
  \end{center}
\end{figure}

We performed the stacking in each group of the sample galaxies
classified by the stellar mass and specific SFR ($SSFR=SFR/M_*$)
derived from the MAGPHYS SED fits (see \S\ref{subsec:sed}). The
$SFR-M_*$ relation of our sample is shown in Figure~\ref{fig:SFR-M}.
The SSFR classification was based on the main-sequence (MS) relation
from Eq.11 of \cite{Boog18}, using the mean redshift of the \cotwoone
line detectable in Band~3 \citep[$z=1.43$;][]{Walt16}.  Galaxies lying
within, above, and below the intrinsic scatter
(0.44\,dex\,\footnote{This intrinsic scatter is found by assuming a
  Gaussian function in a model of the star formation sequence.  As
  noted in \cite{Boog18}, this value is higher than the average value
  reported in previous work.}) of this relation are referred to as the
``MS'', ``above'' the MS, and ``below'' the MS, respectively,
throughout this paper. We here use the main-sequence relation from
\cite{Boog18} because their $SFR-M_*$ correlation is assessed with the
objects whose spectral features and redshifts were measured based on
the MUSE data. Compared with earlier studies such as \cite{Whit14} and
\cite{Schr15} whose samples are mostly massive
galaxies, \cite{Boog18} better constrained the low-mass end of the
relation. The numbers of galaxies in each group for stacking are
presented in Table~\ref{tbl:bins}.

The main-sequence relation of \cite{Boog18} was also adopted for
classifying galaxies which expected to have CO $J>2$ lines. Similar to
\cotwoone, we used the mean redshifts of $z=2.61$, $3.80$, $4.99$, and
$6.18$ for \cothreetwo, CO(4-3), CO(5-4), and CO(6-5) lines,
respectively \citep[][]{Walt16}.

% Fig.3

\subsection{CO line flux and upper limit measurements}\label{subsec:meas}

For each stacked spectrum, we performed a best fit on the 2D image
(moment-0) with a 2D Gaussian function to estimate the line flux or
upper limit.  The x- and y-positions are allowed to vary in the
vicinity of the MUSE positions within a radius of $1/3$ of the ALMA
beam size. The widths were fixed to the beam size.  We considered a CO
emission line to have been detected when the amplitude of the fitted
Gaussian function was higher than $3\sigma$ of the local fluctuations
in the 2D image.  The line flux was estimated by integrating the
fitted Gaussian function.  When the CO line was not detected, the
standard deviation in the central area of the image was used to
evaluate the $3\sigma$ upper limit, assuming a point source.

Uncertainties in MUSE redshifts can cause some flux losses when we
stack spectra. We assumed an emission line with a width of
$300\,{\rm km\,s^{-1}}$ (full width at half maximum, FWHM) and
performed a bootstrap simulation to assess the flux loss resulting
from $\sigma_z = 0.00012(1+z)$ (\S\ref{subsec:muse}).  The total flux
was measured within $\Delta v = 528\,{\rm km\,s^{-1}}$ (corresponding
to 19 slices in frequency; see \S\ref{subsubsec:co21_all} for this
choice of $\Delta v$) centered at the rest-frequency of \cotwoone.
This velocity range was the same as the one we used to create coadded
2D images for flux measurements (see \S\ref{sec:results}).  With 10000
realizations, we found that 92\% of them resulted in less than 3\%
flux loss. We do not scale up our measured fluxes in this work to take
account of this flux loss, because its impact on our line flux
measurements is not significant.

\section{CO emission from stacked spectra} \label{sec:results}

We performed CO emission line stacking from \cotwoone to
CO(6-5). In this section, we will present the resulting stacked
spectra.

\subsection{\cotwoone}\label{subsec:co21}

\subsubsection{Stacking the entire sample}\label{subsubsec:co21_all}

We first performed stacking without binning in stellar mass or SSFR
to obtain a constraint on the average properties of all galaxies.  The
entire sample (\TotNCOTwoOne spectra) and a subset of the sample
(\TotEXNCOTwoOne spectra) that excluded \DetNCOTwoOne objects whose CO
emission were detected by the blind search (\S\ref{subsec:subsamples})
were used to search for signals in the stacks (below the noise
threshold of individual galaxies).

For these two sets of samples, we inspected 2D images (moment-0) that
were coadded between
$230.335 \leq \nu_{\rm rest}/{\rm GHz} \leq 230.741$
($\Delta v = 528\,{\rm km\,s^{-1}}$) centered at the \cotwoone
  rest frequency ($230.538\,{\rm GHz}$) in the corresponding 1D
spectrum.  This width is consistent with $2\sigma$ of the mean line
width of the ASPECS blind CO detections \citep[$308\,{\rm km\,s^{-1}}$
in FWHM;][]{Arav19} to recover 95\% of the emission.

The total line fluxes were obtained by fitting a 2D Gaussian function
in the 2D images. This measure helps to avoid problems such as a
possible slight positional offset between the optical and CO emission.
All of the Gaussian parameters, the peak position, and amplitude were
set to be free parameters, whereas the widths were fixed to the mean
beam size. We obtained total line fluxes (upper limit) of
$0.031 \pm 0.007 \, {\rm Jy\,km\,s^{-1}}$
($<0.012\, {\rm Jy\,km\,s^{-1}}$) for the stacked spectra including
(excluding) the sources with direct \cotwoone detections.

\subsubsection{Stacking in stellar mass and SSFR bins}\label{subsec:M*-SSFR}

We carried out the stacking on both the entire sample and the sample
that excluded the direct CO detections.  To investigate the dependence
of the molecular gas content on fundamental properties of galaxies, we
divided the sample into ranges of stellar mass and SSFR to perform the
stacking. As shown in Figure~\ref{fig:SFR-M}, we set the bins to have
steps of $\log{(M_*/M_\odot)} = 1.0$ over the range
$8.0 < \log{(M_*/M_\odot)} \leq 12.0$ (4~bins) with stellar mass, and
galaxies ``above'' the MS, on the ``MS'', and ``below'' the MS based
on the main-sequence relation at $z = 1.43$ discussed in
\S\ref{subsec:subsamples}. In the redshift range where \cotwoone can
be detected with the ASPECS Band~3 survey ($z=1.0-1.7$), the
main-sequence relation does not evolve significantly.

\begin{figure*}
  \begin{minipage}{.45\textwidth}
    \begin{center}
      \begin{overpic}
        % [grid,
        [width=\textwidth,trim=0 0 0 35,clip]
        {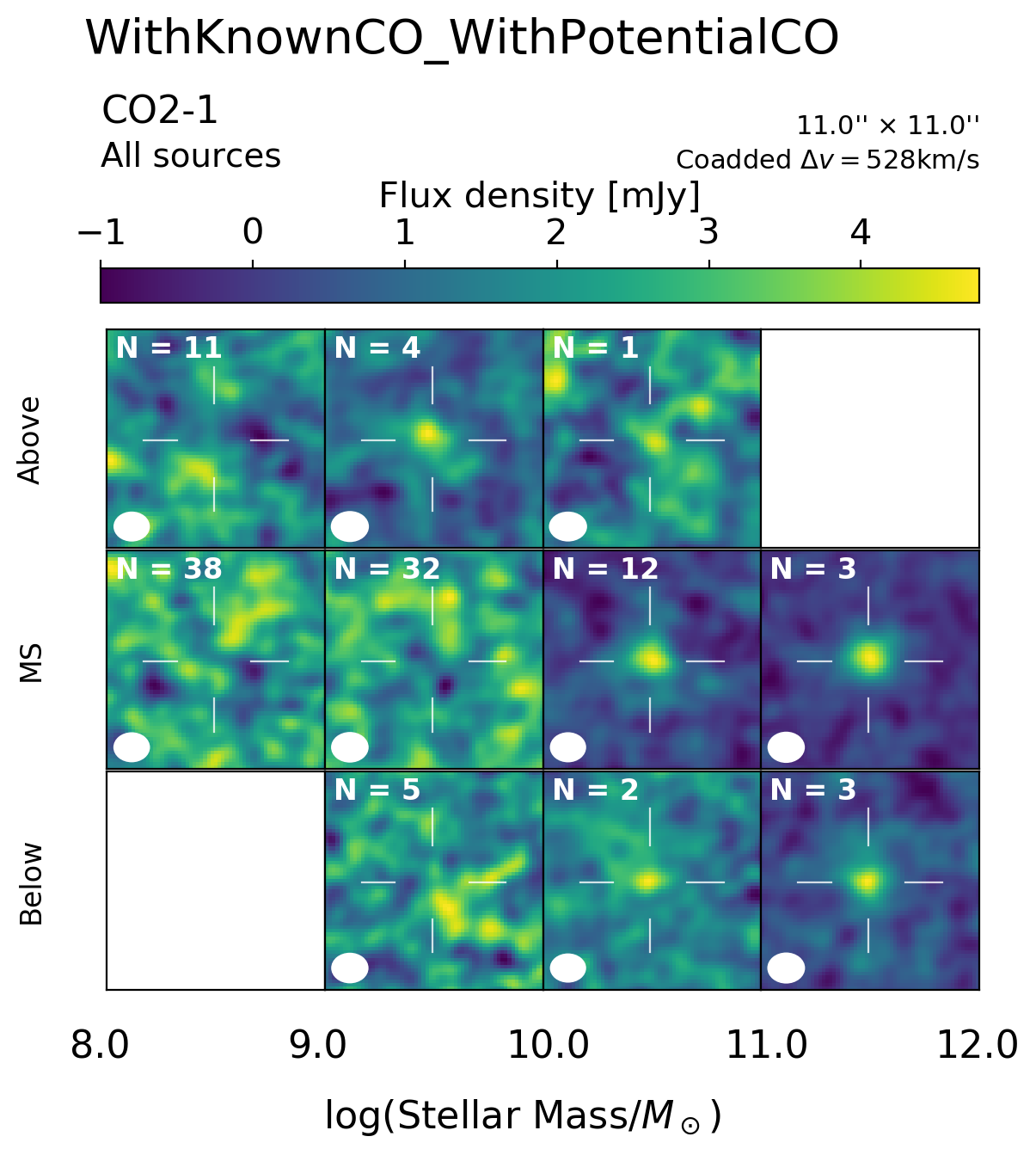}
      \end{overpic}
    \end{center}
  \end{minipage}
%  \hfill
  \begin{minipage}{.50\textwidth}
    \begin{center}
      \begin{overpic}
        % [grid,
        [width=\textwidth,trim=0 0 0 95,clip]
        {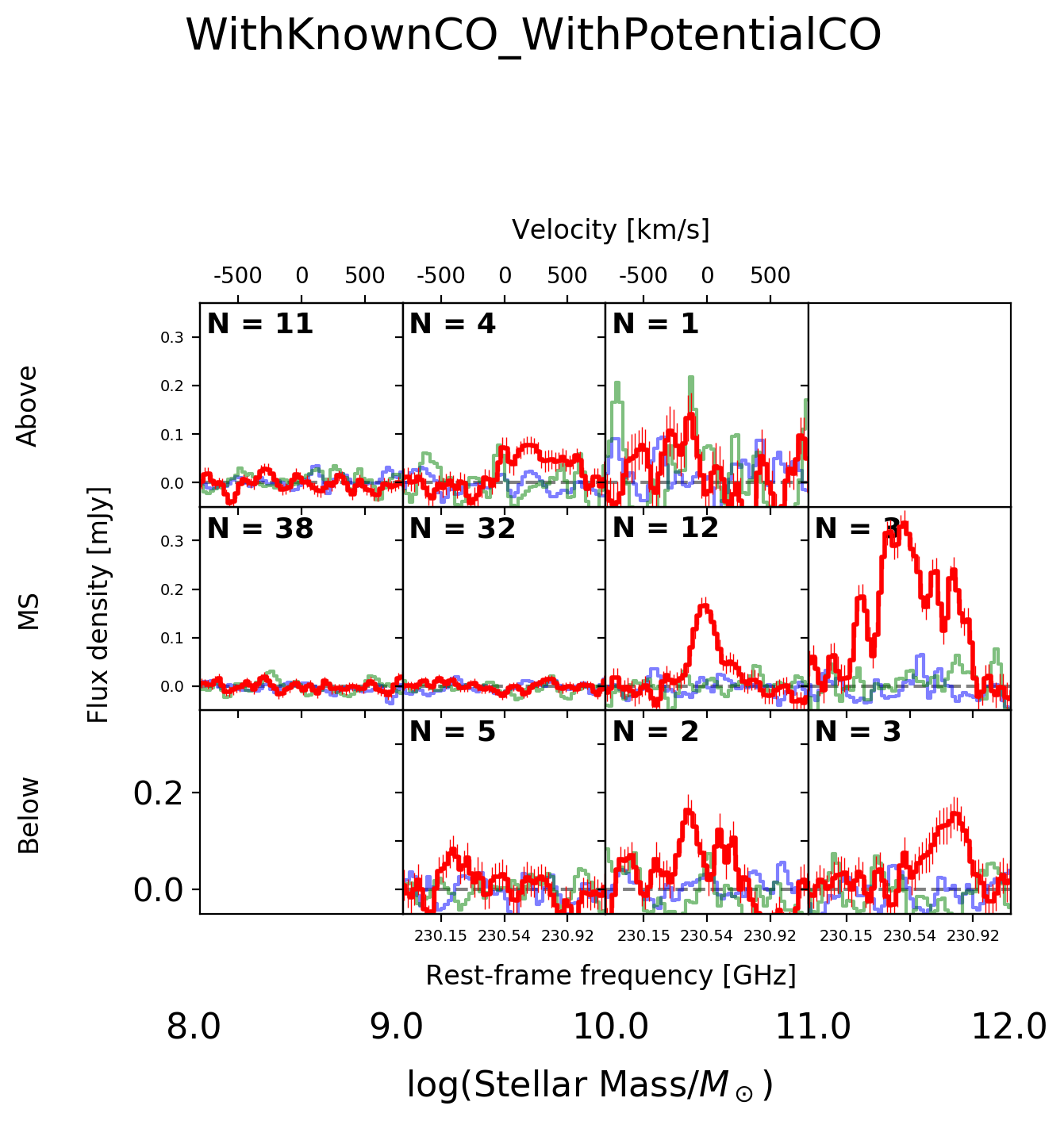}
      \end{overpic}
    \end{center}
  \end{minipage}
  
  \vspace*{1cm}
  
  \begin{minipage}{.45\textwidth}
    \begin{center}
      \begin{overpic}
        % [grid,
        [width=\textwidth,trim=0 0 0 35,clip]
        {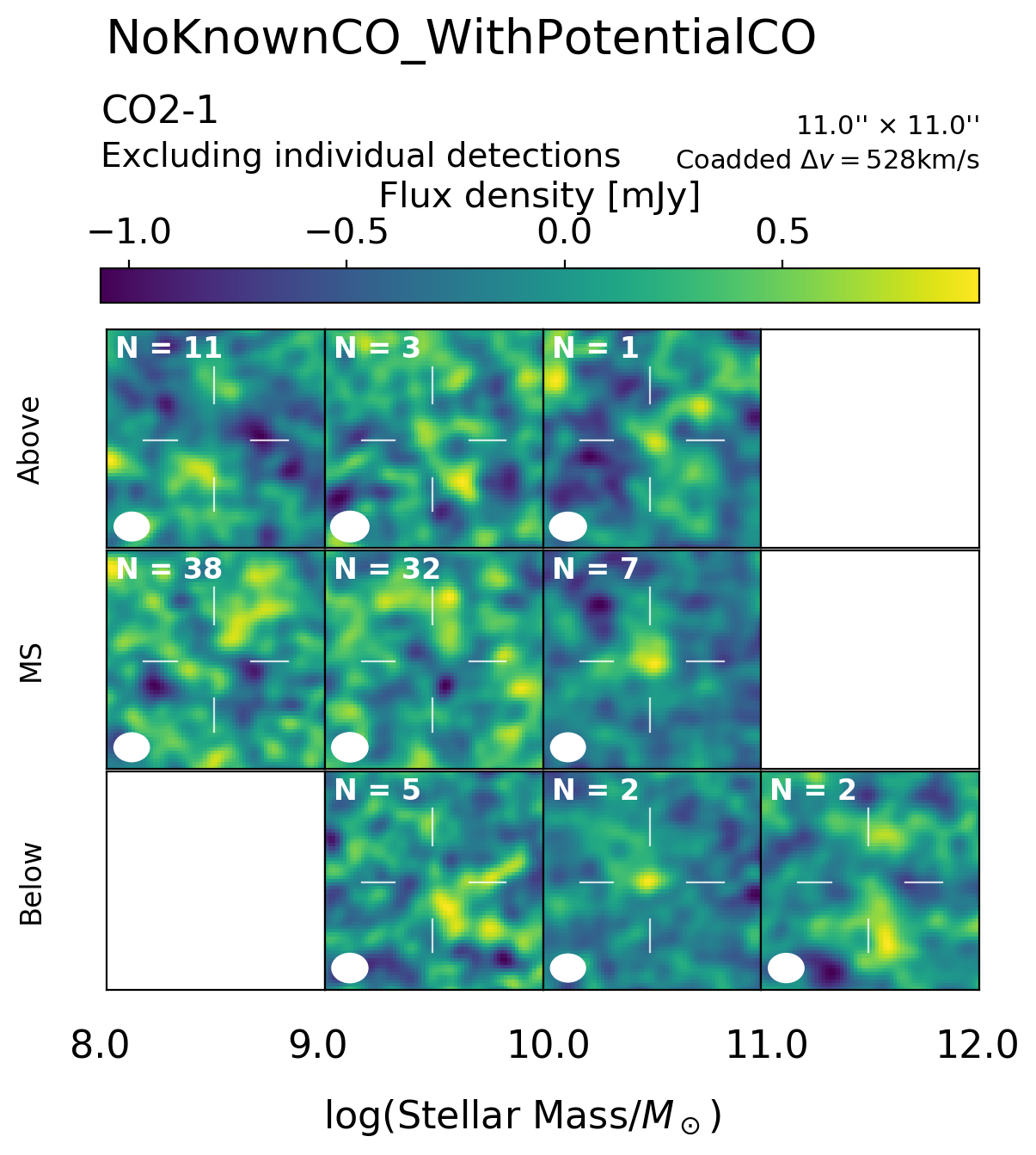}
      \end{overpic}
    \end{center}
  \end{minipage}
%  \hfill
  \begin{minipage}{.50\textwidth}
    \begin{center}
      \begin{overpic}
        % [grid,
        [width=\textwidth,trim=0 0 0 95,clip]
        {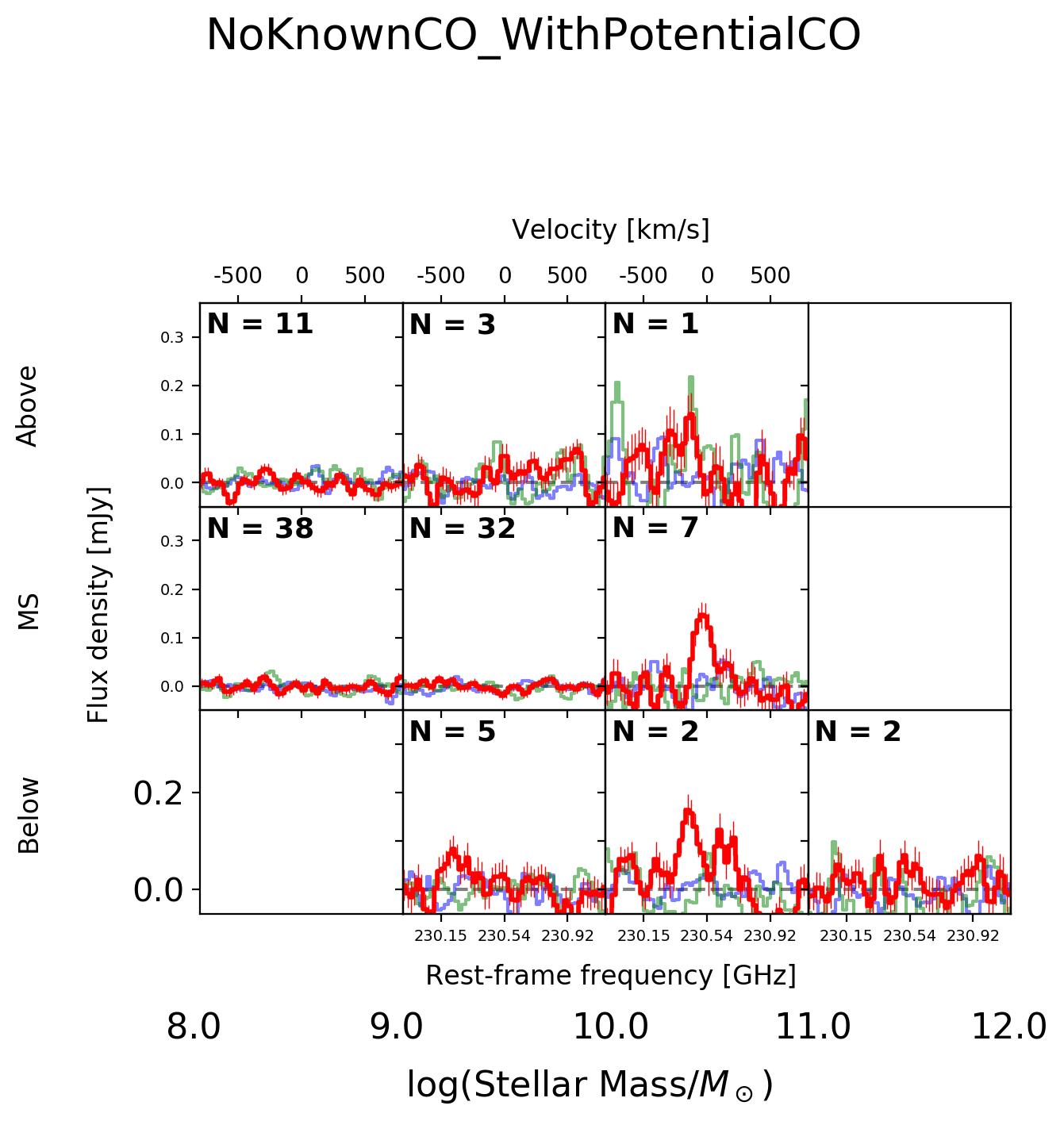}
      \end{overpic}
    \end{center}
  \end{minipage}
  \begin{center}
    \caption{ Resulting stacked 2D images (left) and 1D spectra
      (right) in the bins of SSFRs (above, on, and below the galaxy
      main-sequence, as discussed in \S\ref{subsec:subsamples};
      y-axis) and stellar masses (x-axis).  The upper and lower two
      panels show the results of including and excluding the
      individually reported direct \cotwoone detections, respectively
      \citep{Arav19,Boog19}. The 2D images are coadded between
      $230.335 \leq \nu_{\rm rest}/{\rm GHz} \leq 230.741$
      ($\Delta v = 528\,{\rm km\,s^{-1}}$) and the white ellipses
      indicate the mean of the beam size.  The 1D spectrum is
      extracted from the central spaxel (the red solid line). Spectra
      extracted with random redshifts in place of the MUSE position
      and with random positions at the correct MUSE redshift are shown
      by the light blue and green lines, respectively
      (\S\ref{subsec:extract}).}
  \end{center}
  \label{fig:M8-12_2D1Dstack}
\end{figure*}

The stacked 2D and 1D spectra are shown in
Figure~\ref{fig:M8-12_2D1Dstack} and the CO line flux measurements are
summarized in Table~\ref{tbl:lineflux}.  When the directly detected
\cotwoone emission is included, there are solid detections
($>3\sigma$) at $\log{(M_*/M_\odot)} > 10.0$, regardless of their
SSFRs (except in the $10.0 < \log{(M_*/M_\odot)} \leq 11.0$
bin where there is only a single source). On the other hand, at
$\log{(M_*/M_\odot)} \leq 10.0$, we only obtain a signal in the bin
above the MS with $9.0 < \log{(M_*/M_\odot)} \leq 10.0$.  When we
exclude the known \cotwoone emission from the stacks, a significant
detection is seen in the MS bin with
$10.0 < \log{(M_*/M_\odot)} \leq 11.0$ (seven sources)
\,\footnote{Note that a detection in the bin below the MS with
  $10.0 < \log{(M_*/M_\odot)} \leq 11.0$ has already been observed
  when the known \cotwoone emission were included, because none of the
  galaxies in this bin include the directly detected \cotwoone
  emission (none of the galaxies have been excluded to make any
  changes).}.

Although the bins of the MS galaxies in the ranges of
$8.0 < \log{(M_*/M_\odot)} \leq 9.0$ and
$9.0 < \log{(M_*/M_\odot)} \leq 10.0$ contain the largest numbers of
galaxies, we do not detect a \cotwoone line.  To attempt to detect
stacked CO emission for galaxies in these low mass bins, we adopt a
$\Delta \log{(M_*/M_\odot)}=0.5$ stellar mass bin size in case some
hidden emission from a small number of sources has been diluted by
averaging too many sources without any emission. This smaller bin size
results in 14, 24, 23, and 9 MS sources from
$8.0 < \log{(M_*/M_\odot)} \leq 8.5$ to
$9.5 < \log{(M_*/M_\odot)} \leq 10.0$ in steps of
$\log{(M_*/M_\odot)} = 0.5$.  No detection is identified even with
these finer bins.  The estimated $3\sigma$ upper limits with the 2D
data are $0.028$, $0.024$, and $0.017$,
$0.030 \, {\rm Jy\,km\,s^{-1}}$, from lower to higher stellar mass
bins, respectively.

Among the bins with stacked detections, although the censoring
fraction is unity for the bin of below the MS with stellar mass
$10.0 < \log{(M_*/M_\odot)} \leq 11$, all of the remaining bins are
$\leq 0.75$. In particular, the fraction is $\leq 0.6$ for the MS
galaxies.

% Fig.4

\subsubsection{The \cotwoone lines detected in individual galaxies}\label{subsec:indiv}

We inspect individual spectra of galaxies in the MS bin of
$10 < \log{(M_*/M_\odot)} \leq 11$, where the stacked spectrum has the
highest SNR (the spectra are depicted in Appendix~\ref{app:indiv}).
Excluding the emission already found by the blind search, it is
possible to identify the \cotwoone line with lower SNR in MUSE IDs
879, 985, and 1308. We note that MUSE IDs 879 and 985 have already
been reported by \cite{Boog19} as the MUSE prior-based sample (see
their Table~2 and Figure~4).  A further potential stacked \cotwoone
detection is seen in the range of
$10.0 < \log{(M_*/M_\odot)} \leq 11.0$ for the mean of two
galaxies lying below the MS.
A detected \cotwoone line is dominated by MUSE ID 928.

\subsection{Higher-$J$ CO stacked spectra}\label{subsec:results_hiJ}

Along with the \cotwoone stacking analysis, we attempt to detect
higher excitation CO emission, up to $J=6-5$, with the same stacking
method.  The census of the stacked sources is shown in
Table~\ref{tbl:hi_j}.

For \cothreetwo, among our sample selection
(\S\ref{subsec:subsamples}), MUSE~IDs~35 and 1124\,\footnote{
      This source has a new additional MUSE redshift
      $z=2.5739$ whose foreground galaxy is identified at
      $z=1.098$ \citep{Boog19}.}  are the only sources that
were already known from the blind search (ASPECS-LP-3mm.01 and
  3mm.12). The former is a galaxy above the MS and the latter is
  on the MS with $\log{(M_*/M_\odot)} = 10.39$ and 
  10.64, respectively. In addition, we consider MUSE~ID~35's
pair galaxy MUSE~ID~24 ($\log{(M_*/M_\odot)} = 9.45$) to have a
\cothreetwo detection in our analysis. This is because the CO spatial
extent, which peaks at the location of ID~35, covers ID~24 and it is
difficult to distinguish the flux contributions from these two
sources. When the stacking was performed in the same way as for
\cotwoone, these two sources dominated the detections in their bins
(Appendix~\ref{app:hi_j}). If we discard them, no significant emission
remains. This result agrees with the finding of \cite{Uzgi19} who also
did not find additional \cothreetwo emission with a masked auto-power
spectrum using all MUSE positions with LP primary beam response
$\geq 20\%$.  There is also no stack detection for $J_{\rm up} \ge 4$
CO emission in our sample with the blind search~\footnote{There is one
  CO(4-3) detection in the blind search, but this source does not have
  a MUSE redshift \citep{Arav19,Boog19}.}, nor the stacking.

We note that the MUSE spectroscopic redshifts of galaxies at
$z \gtrsim 3$, which allow detections of $J_{\rm up} > 4$ CO emission,
were mostly determined using the Ly$\alpha$ line.  Its redshift was
measured using the peak emission, which can have a few hundred
$\rm km\,s^{-1}$ offset from the systemic redshift
\citep[e.g.,][]{Shap03}.  When the redshifts of the stacked sources
were measured with Ly$\alpha$, we tried to recover their systemic
redshifts based on an empirical correlation between the Ly$\alpha$
emission peak and the Ly$\alpha$ line width \citep{Verh18}.  The
intrinsic scatter of the relation is
$\pm 73\,{\rm km\,s^{-1}}$ measured with 13 sources that have
both Ly$\alpha$ and \ion{C}{3}] detections.  We do not find any
detection for $J_{\rm up} > 4$ CO emission either. This is likely
either due to dilution of the stacked signal, owing to uncertainties
in the velocity offset, or to Ly$\alpha$ emitters on average having
smaller SFRs and molecular gas content.

We also visually investigate individual CO spectra with the MUSE
redshifts as priors to search for possible CO emission which is washed
out by stacking. Two potential CO(5-4) emission lines were
identified as shown in Figure~\ref{fig:CO5-4} in
Appendix~\ref{app:CO5-4}. If confirmed, they may be one of the first
cases of high-$J$ CO detection at $z > 4.5$ for Ly$\alpha$
emitters. Deeper observations are needed to confirm these
detections.

\section{Discussion} \label{sec:discussion}

In this section, we will concentrate on discussing the molecular gas
content at $z\sim1.5$, as we have obtained the most reliable
measurements at this redshift with our analysis.

\subsection{CO Luminosities and Molecular ($\rm H_2$) Gas Masses} \label{subsec:gas_mass}

We employed the following equation to obtain the CO luminosities from
the measured \cotwoone emission \citep{Solo97,Solo05}:

\begin{equation}
\frac{L^{\rm \prime}_{\rm COJ-[J-1]}}{{\rm K\,km\,s^{-1}\,pc^2}} = 
3.25 \times 10^7 \, 
\frac{F_{\rm line}}{\rm Jy\,km\,s^{-1}} \,
\frac{D_L^2}{(1+z)^3 \nu_{\rm obs}^2}
\end{equation}

\noindent
where $F_{\rm line}$ is the integrated line flux density, $D_L$ is the
luminosity distance in Mpc, and $\nu_{\rm obs}$ is the observed
frequency in GHz.  For the redshift ($z$), we used the mean redshift
of the galaxies in each stacking bin.

We then adopted the following equation and the same conditions as
\cite{Deca16b} to infer molecular gas masses ($M_\odot$) from the line
luminosities (${\rm K\,km\,s^{-1}\,pc^2}$):

\begin{equation}
  M_{\rm gas} = 
  \frac{\alpha_{\rm CO}}{r_{J1}}\, L^{\rm \prime}_{\rm COJ-[J-1]}
\end{equation}

\noindent
where $J$ is the upper level of the CO excitation and
$\alpha_{\rm CO}$ is the CO luminosity to gas mass conversion
factor. We assume $r_{21}=0.76 \pm 0.09$ following \cite{Dadd15}.
Based on galaxies detected with the ASPECS survey, \cite{Boog20}
estimated $r_{21}=0.75 \pm 0.11$, which is comparable to the value
from \cite{Dadd15}. Here we adopt the value from the former to be
consistent with the molecular mass measurements published in
\cite{Arav19}, which are used for comparison in this paper.  For
$\alpha_{\rm CO}$, we used
$3.6 \, M_\odot \, ({\rm K\,km\,s^{-1}\,pc^2})^{-1}$ for galaxies in
the stellar mass bins of $\log{(M_*/M_\odot)} > 10$ which is
one of the best estimates for high-redshift main-sequence star-forming
galaxies \citep{Dadd10b}.  Among the ASPECS directly detected CO
sources with metallicity estimates available from the MUSE spectra,
all have roughly solar metallicity \citep{Boog19}, which further
justifies our choice of $\alpha_{\rm CO}$ \citep{Arav19}.  For the
galaxies with $\log{(M_*/M_\odot)} \le 10.0$, because they
are likely to have lower metallicity, $\alpha_{\rm CO}$ could be
higher \citep[e.g.,][]{Bola13b}.  In this work, we assume that the
metallicities of galaxies with $\log{(M_*/M_\odot)} \sim 9.5$ and
$\sim 8.5$ are $\sim 2/3Z_\odot$ and $\sim 1/2Z_\odot$
\citep[e.g.,][]{Sava05,Erb06,Mann09,Zahi11,Sarg14,Sand20}. Thus,
$\alpha_{\rm CO}$ would increase by a factor of $\sim 2$ and $\sim 3$,
respectively, compared to galaxies with $\log{(M_*/M_\odot)} > 10.0$
\citep[e.g.,][]{Genz12,Schr12}. We adopt $\alpha_{\rm CO} = 7.2$ and
$10.8 \, M_\odot \, ({\rm K\,km\,s^{-1}\,pc^2})^{-1}$ for galaxies
with $\log{(M_*/M_\odot)} \sim 9.5$ and
$\log{(M_*/M_\odot)} \sim 8.5$, respectively. In these lower stellar
mass bins, the $M_{\rm gas}$ estimates with a constant
$\alpha_{\rm CO} = 3.6 \, M_\odot \, ({\rm K\,km\,s^{-1}\,pc^2})^{-1}$
are also shown as dotted symbols in figures for reference. The
calculated line luminosities and molecular gas masses from \cotwoone
are summarized in Table~\ref{tbl:linelum}.

In the following subsections, we discuss the gas content of galaxies
at $z\sim1.5$, together with their stellar masses and SFRs derived
from the SED fitting (\S\ref{subsec:sed}). The \cotwoone emission used
here stems from the measurements based on the 2D Gaussian fits in the
images presented in \S\ref{subsec:M*-SSFR} and
Figure~\ref{fig:M8-12_2D1Dstack}.  Because we are interested in global
properties of the molecular gas, we use the measurements that include
the direct \cotwoone detections from the blind search.

\subsection{Molecular gas and stellar mass scaling relation of MS galaxies}\label{subsec:Mgas_Mstar}

Previous observational and theoretical studies of molecular gas in
high redshift galaxies have established a set of scaling relations
that relate the molecular gas content to galaxy properties such as
stellar masses, SFRs, source sizes
\citep[e.g.,][]{Tacc13,Tacc18,Freu19,Arav19,Genz15,Magd12b,Sant14,
  Scov16,Some12,Dave17,Popp19}. These relations have been key to
understanding how the galaxy growth process has taken place through
cosmic time.  We first compare average molecular gas properties to
the stellar mass of distant galaxies.

In conjunction with the MUSE deep survey in the same field, the
stacking analysis facilitates the exploration of the gas scaling
relations below stellar masses of $\log{(M_*/M_\odot)} = 10$, a regime
that is rather uncharted in CO emission for individual detections at
$z > 1$ (see also Appendix~\ref{app:Spea14}).
We focus on discussing the main-sequence galaxies, which have the
largest numbers of stacked spectra in our \cotwoone sample.

% Fig.6
\begin{figure*}
  \begin{center}
    \includegraphics[angle=0,width=0.48\textwidth]
    {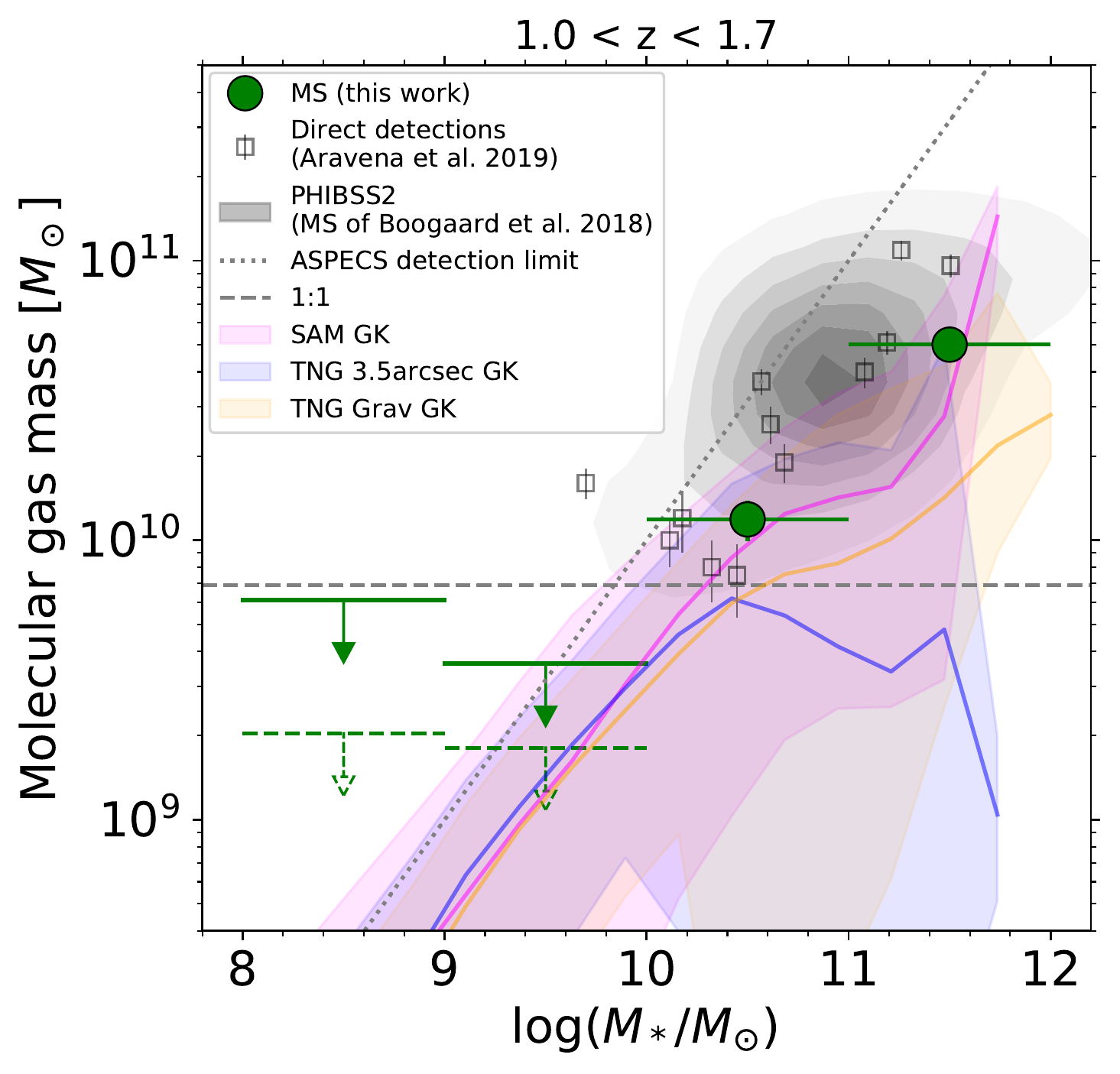}
    \includegraphics[angle=0,width=0.48\textwidth]
    {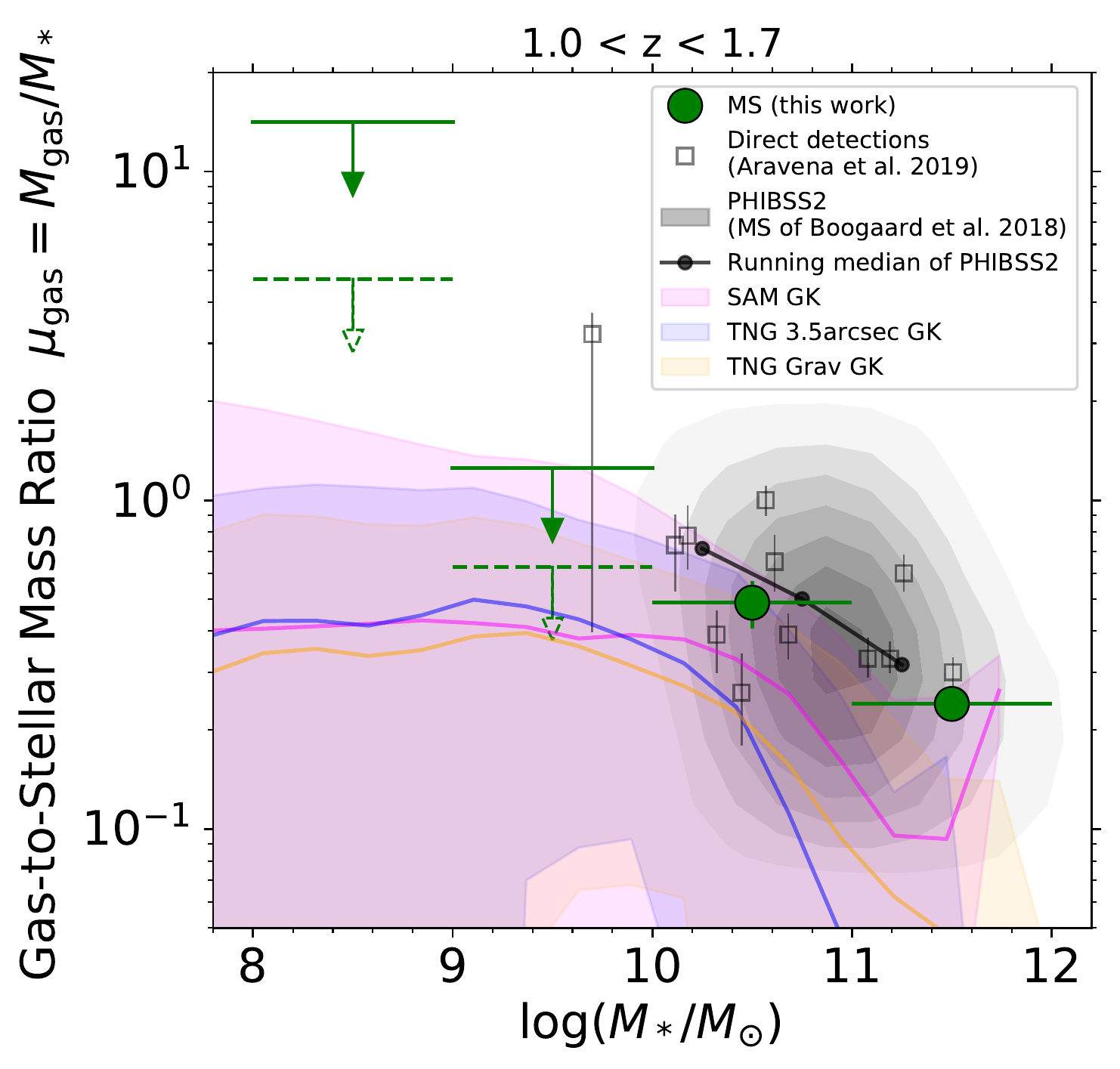}
    \caption{ [Left] Molecular gas mass as a function of stellar mass
      of the main-sequence galaxies in this study. We use
      $\alpha_{\rm CO}=3.6 \, M_\odot \, ({\rm
        K\,km\,s^{-1}\,pc^2})^{-1}$ for galaxies in the
      $\log{(M_*/M_\odot)} > 10$ bins and 7.2 and 10.8 for the
      $\log{(M_*/M_\odot)} \sim 9.5$ and 8.5 bins, respectively, to
      obtain molecular gas mass (see \S\ref{subsec:gas_mass}). We also
      show the estimates based on a constant
      $\alpha_{\rm CO}=3.6 \, M_\odot \, ({\rm
        K\,km\,s^{-1}\,pc^2})^{-1}$ in the lower mass bins with the
      dashed symbols for reference.  The large green circles are from
      the stacked CO detections and the downward-pointing arrows are
      the $3\sigma$ upper limits.  The squares indicate the ASPECS CO
      detections from the blind search \citep{Arav19}. The PHIBSS2
      galaxies \citep{Tacc13,Tacc18,Freu19}, classified as MS galaxies
      with the MS relation of \cite{Boog18} over
      $1.00 \leq z \leq 1.74$, are shown as the gray contours for
      comparison to our results.  We also display the model
      predictions of molecular gas mass from SC SAM (pink line and
      shading, which indicate the median and $2\sigma$ scatter,
      respectively), IllustrisTNG with the 3.5arcsec aperture (blue),
      and IllustrisTNG with the Grav aperture (beige) from
      \cite{Popp19}. [Right] Molecular gas-to-stellar mass ratio is
      presented instead as a function of stellar mass. The symbols are
      the same as in the left panel. The black line with dots shows
      the running median of the gas-to-stellar mass ratios of the
      PHIBSS2 sources lying on the MS of \cite{Boog18} \citep[gray
      contours; see also][]{Tacc18}.}
    \label{fig:Mgas}
  \end{center}
\end{figure*}

In Figure~\ref{fig:Mgas} (left), we plot the molecular gas mass from
the stacking analysis against stellar mass of the main-sequence
galaxies, along with individual ASPECS CO detections \citep{Gonz19,
  Arav19}. The majority of directly detected galaxies have
$\log{(M_*/M_\odot)} > 10$.  At $\log{(M_*/M_\odot)} \leq 10$, there
is only one direct detection.  Although stacking also leads to no
detection, the stacks of 38 and 32 spectra provide tight
constraints on the average gas mass in the range of
$8 < \log{(M_*/M_\odot)} \leq 10$ at $z \sim 1.5$.  Taking upper
limits into account, an increase of gas mass is found with increasing
stellar mass from at least $\log{(M_*/M_\odot)} \sim 9.0$ to $12.0$.

For comparison, we also show the contours of the molecular gas
measurements from the PHIBSS2 survey \citep[included both the CO- and
dust-based measurements;][]{Tacc13,Tacc18,Freu19}.  For further
comparison with the literature, we show the distributions of the
PHIBSS2 galaxies at lower redshifts in Appendix~\ref{app:Spea14}.  The
gas masses of the PHIBSS2 sources were derived via a metallicity-based
prescription for this parameter \citep{Genz12}, resulting in
$\alpha_{\rm CO} \sim 4-7 \, M_\odot \, ({\rm
  K\,km\,s^{-1}\,pc^2})^{-1}$ at $z\sim1.5$.

We also compare our measurements with cosmological galaxy formation
model calculations of molecular gas mass presented in
\cite{Popp19}. In the same diagram we show three different model
predictions: the IllustrisTNG hydrodynamical simulations with the
``3.5arcsec'' and ``Grav'' apertures and the Santa Cruz semi-analytic
model (SC SAM).  The former corresponds to all the $\rm H_2$ within a
radius of $3.5\arcsec$ of the source center (similar to ASPECS) and
the latter corresponds to all the $\rm H_2$ gravitationally bound to
the galaxy.  The $\rm H_2$ properties of galaxies were derived based
on the molecular hydrogen fraction recipe of
\cite{Gned11}\,\footnote{These models are the same as the ones shown in
  Figure~2 in \cite{Popp19}, but without the ASPECS observational
  selection effects (i.e. we show here the entire population of
  galaxies predicted by the models).}.

We use the gas mass model predictions at $z=1.43$ \citep{Popp19},
which is the mean redshift of the \cotwoone line detectable in Band~3.
At the higher stellar mass end, all of the three models predicted
lower gas mass than the observed gas mass. At
$\log{(M_*/M_\odot)} \sim 10.5$, about half of the sources with ASPECS
direct detections lie within the $2\sigma$ scatter of the predictions.
At $\log{(M_*/M_\odot)} \lesssim 10.0$, where the predicted gas mass
is below the ASPECS detection limit, the only CO constraint is from
stacking.  The upper limit at $\log{(M_*/M_\odot)} \sim 9.5$ derived
from the stacking is consistent with the models.  This result also
implies that at least a factor of 10 increase in the sample size
is needed to confirm or rule out the model predictions.

\subsection{Molecular gas-to-stellar mass ratio of MS galaxies as
  a function of stellar mass}\label{subsec:Fgas_Mstar}

We depict this plot with a different presentation in
Figure~\ref{fig:Mgas} (right) to show a more common presentation of
molecular gas mass normalized by the stellar mass (molecular
gas-to-stellar mass ratio, $\mu_{\rm gas} \equiv M_{\rm gas}/M_*$) as
a function of stellar mass.  In the literature, a decrease of the
gas-to-stellar mass ratio with increasing stellar mass at
$\log{(M_*/M_\odot)} > 10.5$ has been reported at $z \sim 1.5$
\citep[e.g.,][]{Tacc13}.  However, no observational CO constraints
exist at lower stellar masses below $10^{10}\,M_\odot$. Our stacking
analysis facilitates exploration of this lower stellar mass regime.

For galaxies with $\log{(M_*/M_\odot)} \sim 10.5$ and $11.5$, we
estimate $\mu_{\rm gas} = $ \MMImTEMS and \MMImETMS, respectively.
Similarly to Figure~\ref{fig:Mgas} (left), these values are in
agreement with the measurements of the PHIBSS2 galaxies lying on the
MS relation of \cite{Boog18} \citep[grey contours; see
Appendix~\ref{app:Spea14} for a version using the MS relation
of][]{Spea14}.  In the lower stellar mass bin,
$\log{(M_*/M_\odot)} \sim 9.5$, we constrain $\mu_{\rm gas}$ to have a
$3\sigma$ upper limit of $<1.25$ with
$\alpha_{\rm CO}=7.2 \, M_\odot \, ({\rm K\,km\,s^{-1}\,pc^2})^{-1}$
(or $< 0.63$ if
$\alpha_{\rm CO} = 3.6 \, M_\odot \, ({\rm
  K\,km\,s^{-1}\,pc^2})^{-1}$).

Beyond $\log{(M_*/M_\odot)} \sim 10.5$, a {\em decrease} of the
gas-to-stellar mass ratio is discerned with increasing stellar mass in
our stack results, as well as in previous work reporting both CO-based
and dust-based gas estimates
\citep{Magd12b,Tacc13,Tacc18,Genz15,Scov17b,Arav19,Liu19}.  
As an example, we show in Figure~\ref{fig:Mgas} the result from
PHIBSS2 with a steady decrease of $\mu_{\rm gas}$ with an increase of
stellar mass (black line).

This decline of $\mu_{\rm gas}$ does not seem to be as
  steep at $\log{(M_*/M_\odot)} < 10.0$. If a constant
  $\alpha_{\rm CO}$ is also adopted for these lower mass bins, the
gas-to-stellar mass ratio is consistent with being constant from
$\log{(M_*/M_\odot)} \sim 9.5$ to $10.5$, then turning down around
$\log{(M_*/M_\odot)} \sim 10.5$.  We note that the models from
\cite{Popp19}, despite a discrepancy with the observed results at the
high mass side, show a similar constant gas-to-stellar mass ratio up
to $\log{(M_*/M_\odot)} \sim 10.5$. This is consistent with our
finding if we assume a constant conversion factor.

Given the limitations of our data, and the unknown dependence of the
$\alpha_{\rm CO}$ conversion factor, we cannot determine at which
stellar mass a possible downturn of the gas-to-stellar mass ratio
occurs.  We, however, note that such a ``plateau'' in the low stellar
mass regime has also been identified for local galaxies with direct
detections of CO emission: here $\mu_{\rm gas}$ stays constant from at
least $\log{(M_*/M_\odot)} \sim 9.0$ and starts decreasing around
$\log{(M_*/M_\odot)} \sim 10.5$ \citep[e.g.,][]{Sain17, Both14}.  A
similar result is found in \cite{Tacc18} who compared local
star-forming galaxies, where CO measurements for low mass galaxies are
available \citep[including][]{Sain17}, to distant galaxies after
removing assumed redshift effects on their gas mass content (see also
Appendix~\ref{app:Spea14}).

The declining $\mu_{\rm gas}$ at the high stellar mass end can be
attributed to stellar feedback
\citep[e.g.,][]{Tacc13,Genz15,Dave11}. In contrast, the flatter
$\mu_{\rm gas}$ at $\log{(M_*/M_\odot)} \lesssim 10.5$ may imply that
the effects of feedback are weaker.  The drop in gas-to-stellar mass
ratio seems to appear around $\log{(M_*/M_\odot)} \sim 10.5$, which is
relevant to the characteristic stellar mass of star-forming galaxies
\citep[e.g.,][]{Dunc14} where mass-quenching is becoming dominant
\citep{Peng10}.

Furthermore, the gas-to-stellar mass ratio below $\sim 1$ at
$3\sigma$ of low stellar mass galaxies suggests that we may have
retrieved most of the CO emission at $z \sim 1.5$.  As shown in
\cite{Deca19}, CO luminosity functions derived from the ASPECS data
are assumed to have a fixed faint-end slope, which is consistent with
the faint-end slope of the stellar mass function of star-forming
galaxies. Thus, the value of $\mu_{\rm gas} \lesssim 1$
indicates that the assumed faint-end slope of the CO luminosity
function is at least consistent or could be flatter than that of
the stellar mass function.  If this is the case, most of the CO
emission at $z \sim 1.5$ has been recovered \citep[see also][]{Uzgi19}.

The estimated molecular gas density $\rho{\rm (H_2)}$ at $z\sim1.5$,
based on our \cotwoone stacking measurements, is
$(0.49 \pm 0.09) \times 10^8 \,{\rm M_\odot\,Mpc^{-3}}$.  This value
is lower than, but formally consistent with, the value derived
in \cite{Deca19} with a different CO selection. The CO emission used
in \cite{Deca19} included sources that did not enter the present
analysis because the lack of a counterpart with high-quality
MUSE redshift.  Our total CO flux estimate from stacking is also in
line with the result from the CO auto-power spectrum analysis using
the MUSE positions \citep{Uzgi19}.

\subsection{Dependence of the molecular gas content on SSFR}

In the previous section we only considered main-sequence galaxies; in
the remaining discussion, we will include galaxies above and below the
main-sequence relation of \cite{Boog18} to discuss the dependence of
molecular gas content on SSFR.  Below, we keep the same assumed
$\alpha_{\rm CO}$ conversion factors as above: 3.6, 7.2, and
$10.8 \, M_\odot \, ({\rm K\,km\,s^{-1}\,pc^2})^{-1}$ for galaxies
with $\log{(M_*/M_\odot)} > 10$, $\sim 9.5$, and $8.5$,
respectively (\S\ref{subsec:gas_mass}). We include the constraints
using a constant $\alpha_{\rm CO}=3.6$ throughout the stellar mass
bins in the figures for reference.

\subsubsection{Gas-to-stellar mass ratios}\label{subsec:gas_frac}

\begin{figure*}
  \begin{center}
    \includegraphics[angle=0,width=0.49\textwidth]
    {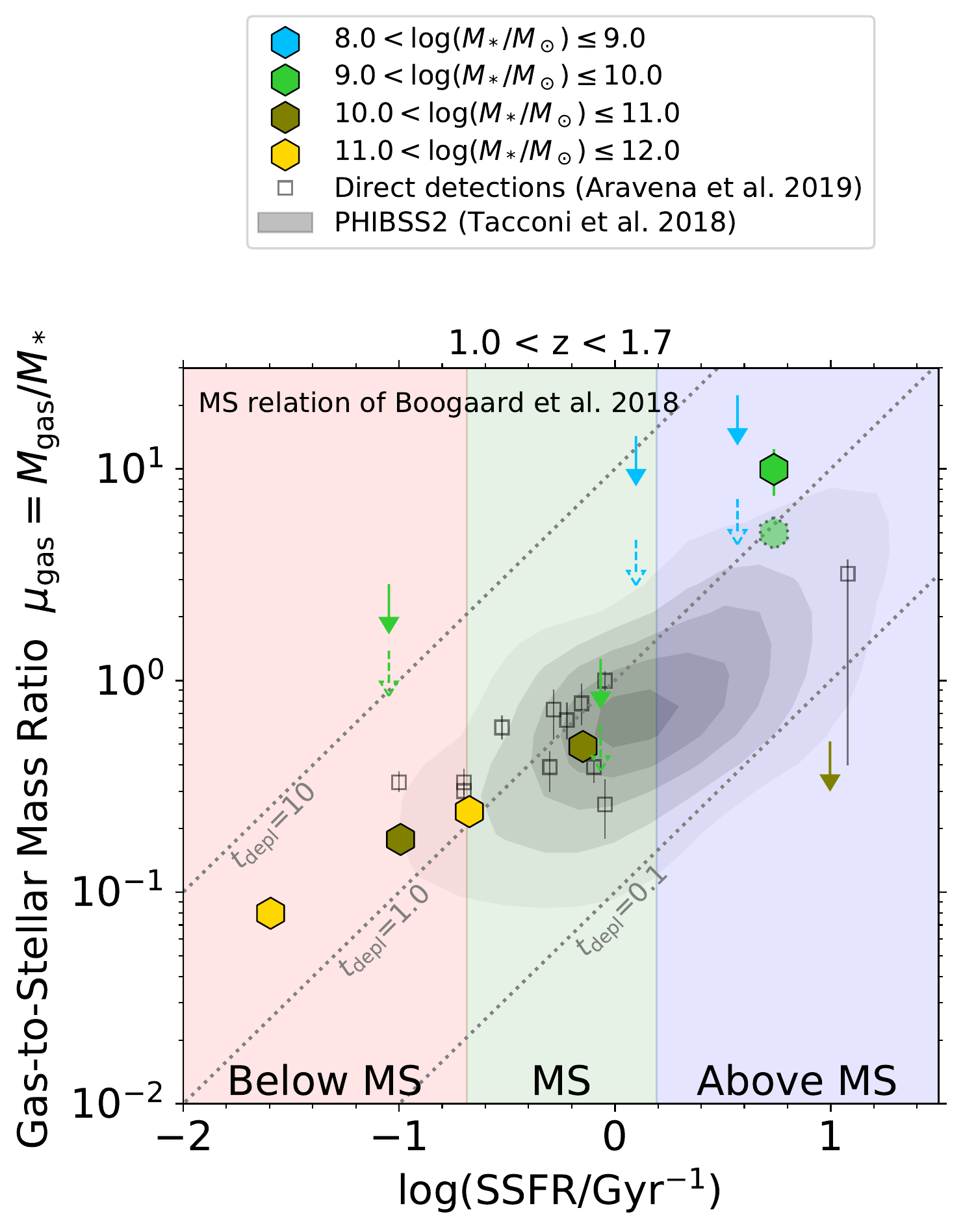}
    \includegraphics[angle=0,width=0.49\textwidth]
    {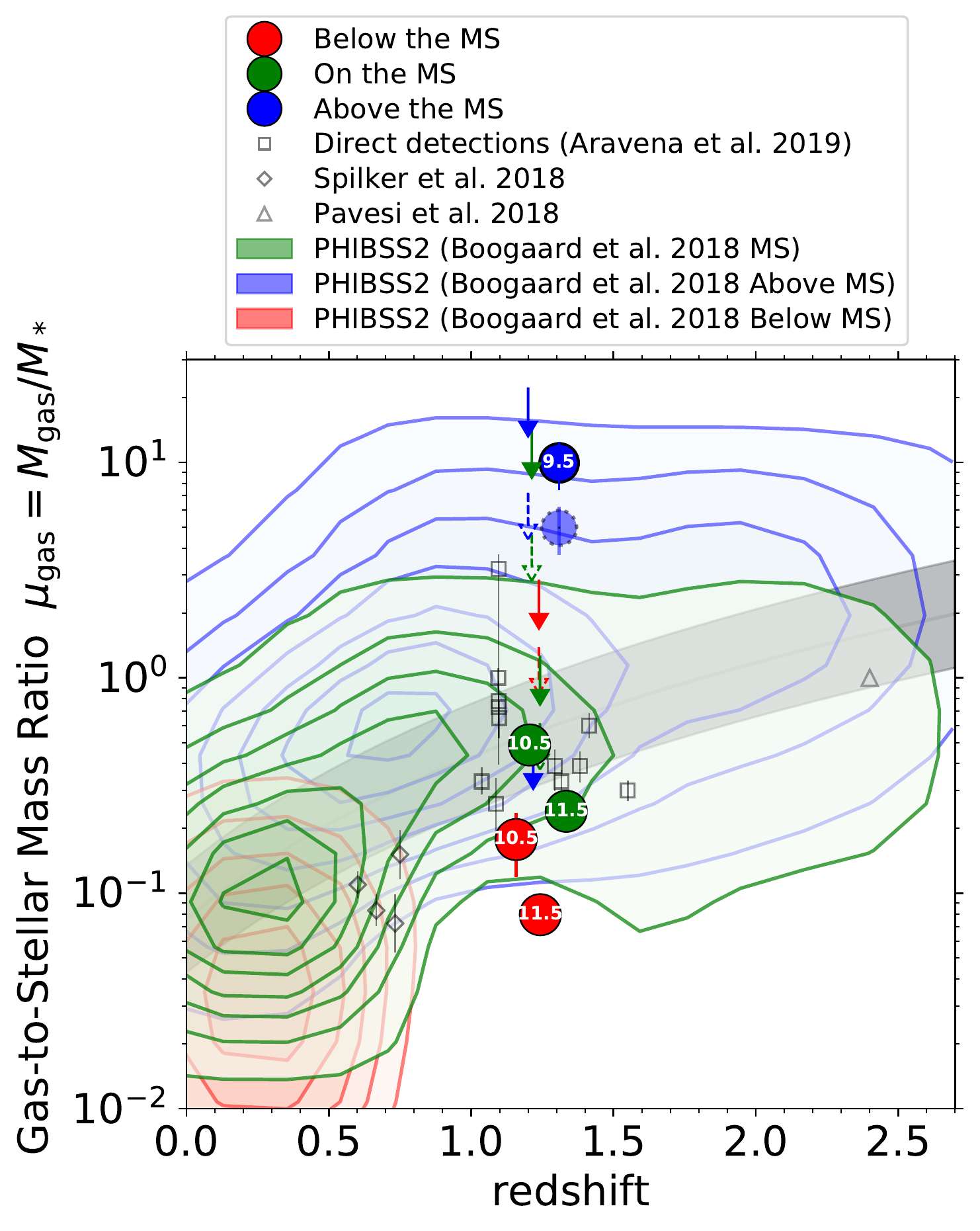}
    \caption{ [Left] Gas-to-stellar mass ratios as a function of
      SSFR. The symbols are the same as in Figure~\ref{fig:Mgas}, but
      now we show the entire sample, below, on, and above the
      main-sequence relation, color-coded with their stellar masses as
      indicated in the legend.
      The same as the other figures involving molecular gas mass
      estimates, we use
      $\alpha_{\rm CO}=3.6 \, M_\odot \, ({\rm
        K\,km\,s^{-1}\,pc^2})^{-1}$ for galaxies in the
      $\log{(M_*/M_\odot)} > 10$ bins and 7.2 and 10.8 for
      the $\log{(M_*/M_\odot)} \sim 9.5$ and 8.5 bins,
      respectively (see \S\ref{subsec:gas_mass}). We also show the
      estimates based on a constant
      $\alpha_{\rm CO}=3.6 \, M_\odot \, ({\rm
          K\,km\,s^{-1}\,pc^2})^{-1}$ in the lower mass bins with the
      dashed symbols for reference.  The data points are centered on
      the mean values of the parameters on the x-axis.  The three
      diagonal gray dotted lines represent constant depletion time
      scales of 0.1, 1.0, and 10\,Gyr from the bottom to the top. The
      background pink, green, and blue filled colors indicate the
      below, on, and above the main-sequence relation of \cite{Boog18}
      for a galaxy at $\log{(M_*/M_\odot)} = 10.0$.  A version using
      the main-sequence relation of \cite{Spea14} is shown in
      Appendix~\ref{app:Spea14}.  [Right] The evolution of
      gas-to-stellar mass ratios. The symbols are the same as
      Figure~\ref{fig:Mgas} but now we include galaxies above and
      below the main-sequence relation: above (blue), below (red), and
      on the MS relation (green). The values inside each symbol
      indicate the corresponding stellar mass bins in log scale.  The
      PHIBSS2 galaxies classified as above, below, and on the MS by
      the main-sequence relation of \cite{Boog18} are also depicted as
      red, green, and blue contours, respectively. The gray curve in
      the background and its shaded region show the evolutionary track
      of main-sequence galaxies from \cite{Tacc18}.}
    \label{fig:fgas-SSFR}
  \end{center}
\end{figure*}

As shown earlier, the gas-to-stellar mass ratio ($\mu_{\rm gas}$)
provides information on the supply and depletion of gas reservoirs in
galaxies.  We compare $\mu_{\rm gas}$ against SSFRs in the left panel
of Figure~\ref{fig:fgas-SSFR}.
We find that $\mu_{\rm gas}$ increases
with increasing SSFRs, which is also seen in earlier studies of
high-redshift galaxies, including the dust-based measurements of the
molecular gas mass \citep[e.g.,][]{Tacc13, Tacc18, Genz15, Magd12b,
  Scov16, Liu19}.  With the stacking analysis, we show that this trend
holds even for galaxies below the main-sequence at $z \sim 1.5$.

% Fig.7

In the right panel of Figure~\ref{fig:fgas-SSFR}, we show
$\mu_{\rm gas}$ of stacked sources color-coded by different SSFRs to
compare with their evolution \citep{Geac11, Magd12b}. Because our data
points are derived from the \cotwoone line, all of them lie between
$z=1.0-1.5$. A wide spread is related to the variations of
$\mu_{\rm gas}$ across the main-sequence relation seen in the left
panel \citep[see also e.g.,][]{Tacc18}.  The lower $\mu_{\rm gas}$ of
low SSFR sources is also found by \cite{Spil18} who investigated
galaxies lying below the MS with $\log{(M_{\rm gas}/M_\odot)} \sim 11$
at $z \sim 0.7$.  The depletion times of our sample below the MS are
on average comparable with Spilker et al.'s sample.  Following
  the same assumption as \cite{Spil18}, if these galaxies continuously
  consume the existing gas with the observed SFR,
  then the $\mu_{\rm gas}$ of the galaxies with
  $\log{(M_{\rm gas}/M_\odot)} \sim 10.5$ (11.5) would reduce
  to $1/10$ at $z \approx 0.5$ (0.2) and
  $1/100$ at $z \approx 0.1$ (0.0). Hence, by
$z=0$ their $\mu_{\rm gas}$ would be comparable to those of passive
galaxies at the current epoch.

\subsubsection{Gas depletion time}\label{subsec:T_depl}

\begin{figure}
  \begin{center}
    \includegraphics[angle=0,width=0.5\textwidth]
    {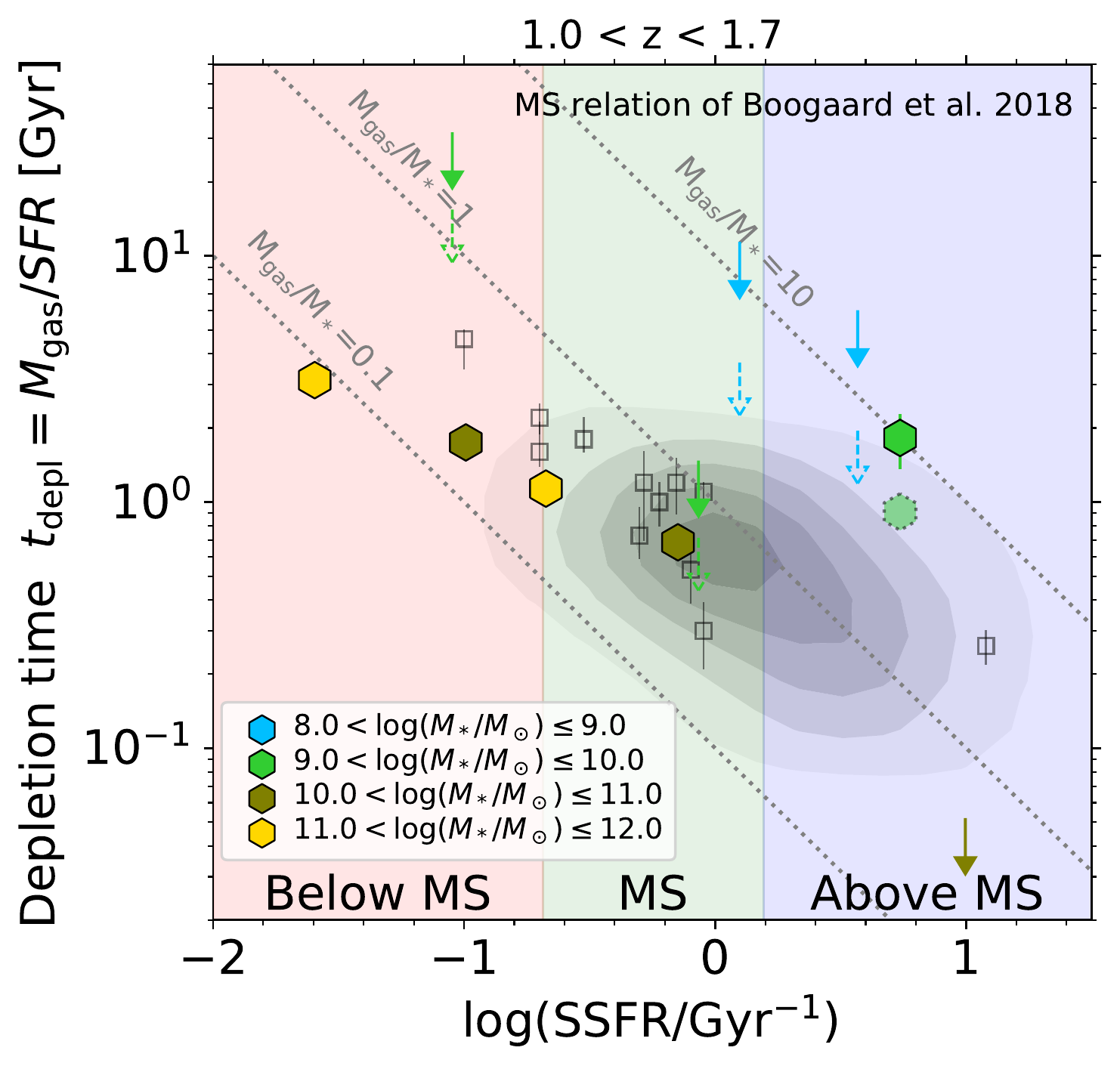}
    \caption{ Depletion time ($t_{\rm depl}$) plotted against
      SSFR. The symbols are the same as Figure~\ref{fig:fgas-SSFR}
      (left).  The same as the earlier figures, the molecular gas
        masses are estimated based on
        $\alpha_{\rm CO}=3.6 \, M_\odot \, ({\rm
            K\,km\,s^{-1}\,pc^2})^{-1}$ for galaxies in the
        $\log{(M_*/M_\odot)} > 10$ bins and 7.2 and 10.8 for
        the $\log{(M_*/M_\odot)} \sim 9.5$ and 8.5 bins,
        respectively (see \S\ref{subsec:gas_mass}). We also show the
        estimates based on a constant
        $\alpha_{\rm CO}=3.6 \, M_\odot \, ({\rm
            K\,km\,s^{-1}\,pc^2})^{-1}$ in the lower mass bins with
        the dashed symbols for reference.  The colors of the filled
      symbols, shown in the legend, indicate each mass bin used for
      the stacking. The three diagonal gray dotted lines represent
      constant gas-to-stellar mass ratios of 0.1, 1.0, and 10 from the
      bottom to the top.}
    \label{fig:Tdepl-SSFR}
  \end{center}
\end{figure}

The gas depletion time, $t_{\rm depl} = M_{\rm gas}/SFR$, is another
way of examining the gas content in galaxies. It estimates the time
taken for the gas to be fully consumed at the current SFR without
accounting for additional fueling.  We show the gas depletion time as
a function of SSFR in Figure~\ref{fig:Tdepl-SSFR}.  Our results from
the stacking analysis are in good agreement with previous
studies. Both the stacked and directly detected sources show
decreasing $t_{\rm depl}$ with increasing SSFR, except that the
  $9.0 < \log{(M_*/M_\odot)} \le 10.0$ bin above the MS shows
  an elevated $t_{\rm depl}$ value.  The data points from these two
sets of measurements occupy the same range in the diagram, following
the constant gas-to-stellar mass ratio ($M_{\rm gas}/M_*$) from around
0.1 to 1.0.  The decreasing gas depletion timescale with SSFR
demonstrates that galaxies with more extreme star formation
consume their gas at a higher rate.  This may be the major cause of
the scatter as shown in Figure~\ref{fig:fgas-SSFR}.

%Fig.8

\section{Summary and Conclusions} \label{sec:summary}

Based on the accurate redshifts from the MUSE IFU survey in the Hubble
Ultra Deep Field, we perform a CO emission stacking analysis with the
ASPECS Band~3 data. When we split the sample into stellar mass and
SSFR bins (on, below, and above the main-sequence relation), we detect
\cotwoone emission ($z\sim1.43$) down to $\log{(M_*/M_\odot)} = 10.0$,
even after removing previously reported CO detections. We do not
recover any higher-$J$ CO emission at higher redshift ($z \gtrsim 3$)
with stacking when excluding the sources with direct CO detections.

The $3\sigma$ upper limits on \cotwoone emission at
$\log{(M_*/M_\odot)} < 10.0$ in the redshift range $1.0 < z < 1.7$
provide meaningful upper limits on molecular gas mass estimates of
main-sequence star-forming galaxies in this stellar mass range, which
has been poorly explored at $z \sim 1.5$.  Under the assumption of
a metallicity-based $\alpha_{\rm CO}$ conversion
  factor, we observe an increase in gas mass with increasing stellar
mass from $\log{(M_*/M_\odot)} \sim 9.0$ to $11.0$. The upper
  limits at the low mass end are consistent with the model
  predictions, but to confirm or rule out these predictions, at least
  a factor of 10 increase in the sample size is needed. The
  gas-to-stellar mass ratio ($\mu_{\rm gas} \equiv M_{\rm gas}/M_*$)
  from $\log{(M_*/M_\odot)} \sim 9.0$ to $10.5$ declines at a slower
  rate compared with the known decrease of $\mu_{\rm gas}$ at higher
  stellar masses.  If a fixed
  $\alpha_{\rm CO}=3.6 \, M_\odot \, ({\rm
      K\,km\,s^{-1}\,pc^2})^{-1}$ conversion factor is assumed across
  the stellar mass range explored here, $\mu_{\rm gas}$ is consistent
with being constant from at least $\log{(M_*/M_\odot)} \sim 9.0$ to
$10.5$, in agreement with predictions by models and observations at
lower redshifts.

Furthermore, the gas-to-stellar mass ratios of $\sim 0.5$
and $\lesssim 1$ at the stellar masses of
  $\log{(M_*/M_\odot)} \sim 10.5$ and
  $\log{(M_*/M_\odot)} \sim 9.5$, respectively, imply
that the faint end slope of the CO luminosity function is at
  least consistent or could be flatter than the stellar mass
function.  We have successfully recovered the majority of the CO
emission at $z \sim 1.5$ with the stacking analysis.  The molecular
gas density
$\rho{\rm (H_2)}=(0.49 \pm 0.09) \times 10^8 \,{\rm
  M_\odot\,Mpc^{-3}}$ from this stacking analysis is comparable with
the one inferred from a CO-driven selection \citep{Deca19}.

When we compare the gas-to-stellar mass ratio ($\mu_{\rm gas}$)
against SSFR, we confirmed the known correlations to also hold for
galaxies with low SSFRs at $z \sim 1.2$.  The scatter in the
$\mu_{\rm gas}-$SSFR correlation seems to be related to the decrease
of $\mu_{\rm gas}$ at $\log{(M_*/M_\odot)} \gtrsim 10.5$ for
star-forming galaxies.  We also show that the gas-to-stellar mass
ratios of massive galaxies
  ($\log{(M_*/M_\odot)} \sim 11$) below the MS at $z\sim1.2$
are comparable to those at $z \sim 0.7$ \citep{Spil18,Tacc18}.

Our stacking analysis of the combined volumetric surveys of ALMA and
MUSE has let us explore regimes that were uncharted before and that
will remain challenging for investigations that rely on direct CO
detections.

%@arxiver{{muse_z_aspecs_2d_CO2-1_pb0.5_confid2_NoKnownCO_WithPotentialCO}.png,{muse_z_aspecs_2d_CO2-1_pb0.5_confid2_WithKnownCO_WithPotentialCO__Mgas_vs_Mstar_MS_B18-MSonly}.pdf,{muse_z_aspecs_2d_CO2-1_pb0.5_confid2_WithKnownCO_WithPotentialCO__Fgas_vs_Mstar_MS_B18-MSonly}.pdf}

\acknowledgments

The authors would like to thank the referee whose constructive
comments helped improving the manuscript.  The authors thank Ian Smail
for useful suggestions to improve this study. This work was supported
by JSPS KAKENHI Grant Number JP19K23462 (HI). Este trabajo cont \'o
con el apoyo de CONICYT + PCI + INSTITUTO MAX PLANCK DE ASTRONOMIA
MPG190030. FW acknowledges support by ERC Advanced Grant 740246
(Cosmic Gas).  T.D-S. acknowledges support from the CASSACA and
CONICYT fund CAS-CONICYT Call 2018.  D.R. acknowledges support from
the National Science Foundation under grant numbers AST-1614213 and
AST-1910107 and from the Alexander von Humboldt Foundation through a
Humboldt Research Fellowship for Experienced Researchers. MK
acknowledges support from the International Max Planck Research School
for Astronomy and Cosmic Physics at Heidelberg University (IMPRS-HD).
ALMA is a partnership of ESO (representing its member states), NSF
(USA) and NINS (Japan), together with NRC (Canada), NSC and ASIAA
(Taiwan), and KASI (Republic of Korea), in cooperation with the
Republic of Chile. The Joint ALMA Observatory is operated by ESO,
AUI/NRAO and NAOJ.


\vspace{5mm}
\facilities{ALMA, VLT:Yepun}


\software{
  astropy \citep{AsPy13,AsPy18},
  CASA \citep{McMu07}
}



\clearpage

%%%%%%%%%%%%%%%%%%%%%%%%%%%%%%%%%%%%%%%%%%%%%%%%%%%
% Figures
%%%%%%%%%%%%%%%%%%%%%%%%%%%%%%%%%%%%%%%%%%%%%%%%%%%

% Tbl1.
\begin{deluxetable}{ccccc}
\tablecaption{Numbers of galaxies in the stellar mass and SSFR bins for
the \cotwoone stacking sample \label{tbl:bins}}
\tablehead{
\colhead{} & \multicolumn4c{$\log{(M_*/M_\odot)}$} \\
\colhead{} & \colhead{$8.0-9.0$} & \colhead{$9.0-10.0$} &
             \colhead{$10.0-11.0$} & \colhead{$11.0-12.0$}
}
\startdata
Above the MS & 11 (11) &  4  (3) &  1 (1) & 0 (0) \\
   On the MS & 38 (38) & 32 (32) & 12 (7) & 3 (0) \\
Below the MS &  0  (0) &  5  (5) &  2 (2) & 3 (2) \\
\enddata
\tablecomments{The numbers in parentheses are after excluding the
  galaxies which have the \cotwoone line identified by the blind search
  (see \S\ref{subsec:subsamples}).}
\end{deluxetable}

% Tbl.2
\begin{deluxetable*}{rcccc}
\tablecaption{Summary of measured stacked \cotwoone line fluxes
  ($F_{\rm line}$) \label{tbl:lineflux}}
\tablehead{
\colhead{} & \multicolumn4c{$\log{(M_*/M_\odot)}$} \\
\colhead{} & \colhead{$8.0-9.0$} & \colhead{$9.0-10.0$} &
             \colhead{$10.0-11.0$} & \colhead{$11.0-12.0$}
}
\startdata
Above the MS & \FImENSB (\FEmENSB) & \FImNTSB (\FEmNTSB) & \FImTESB (\FEmTESB) & \FImETSB (\FEmETSB) \\
   On the MS & \FImENMS (\FEmENMS) & \FImNTMS (\FEmNTMS) & \FImTEMS (\FEmTEMS) & \FImETMS (\FEmETMS) \\
Below the MS & \FImENQU (\FEmENQU) & \FImNTQU (\FEmNTQU) & \FImTEQU (\FEmTEQU) & \FImETQU (\FEmETQU) \\
\enddata
\tablecomments{ The units are in ${\rm Jy\,km\,s^{-1}}$. Values in
  parentheses are after excluding the galaxies which have the
  \cotwoone line identified by the blind search (see
  \S\ref{subsec:subsamples}).}
\end{deluxetable*}

% Tbl.3
\begin{deluxetable*}{rcccccc}
\tablecaption{Total number of galaxies used for stacking high-$J$ CO emission}
\tablehead{
  \colhead{} & \multicolumn5c{$\log{(M_*/M_\odot)}$} \\
  \colhead{} & \colhead{CO} &
   \colhead{$7.0-8.0$} &   \colhead{$8.0-9.0$} &
  \colhead{$9.0-10.0$} & \colhead{$10.0-11.0$} & \colhead{$11.0-12.0$} }
\startdata
\multirow{4}{*}{Above the MS} & 3-2 &  1  (1) &  5  (5) &  5  (4) &  1  (-) &     - \\
                              & 4-3 &  4  (4) &  4  (4) &  5  (5) &  1  (1) &     - \\
                              & 5-4 &       - &       - &       - &       - &     - \\
                              & 6-5 &       - &       - &       - &       - &     - \\
\hline                              
   \multirow{4}{*}{On the MS} & 3-2 &  4  (4) & 21 (21) &  9  (9) &  4  (3) &     - \\
                              & 4-3 & 63 (63) & 79 (79) & 29 (29) &  4  (4) &     - \\
                              & 5-4 & 40 (40) & 46 (46) & 26 (26) & 11 (11) & 2 (2) \\
                              & 6-5 &  2 ( 2) & 16 (16) & 14 (14) &  7  (7) & 3 (3) \\
\hline                              
\multirow{4}{*}{Below the MS} & 3-2 &      -  &       - &       - &       - &     - \\
                              & 4-3 &      -  &       - &  2  (2) &       - &     - \\
                              & 5-4 &  1  (1) &       - &  1  (1) &  1  (1) &     - \\
                              & 6-5 &  2  (2) &       - &       - &       - &     - \\
\enddata
\tablecomments{The numbers in parentheses are after excluding the
  galaxies which have the CO line identified by the blind search
  (see \S\ref{subsec:subsamples}).}
\label{tbl:hi_j}
\end{deluxetable*}

% Tbl.4
\begin{deluxetable*}{rccccc}
  \tablecaption{Summary of \cotwoone line luminosities
    ($L^{\rm \prime}_{\rm CO2-1}$) and molecular gas masses
    ($M_{\rm gas}$) }
  \tablehead{
    \colhead{} & \colhead{} & \multicolumn4c{$\log{(M_*/M_\odot)}$} \\
    \colhead{} & \colhead{} & \colhead{$8.0-9.0$} & \colhead{$9.0-10.0$} &
    \colhead{$10.0-11.0$} & \colhead{$11.0-12.0$} }
  \startdata
  \multirow{2}{*}{Above the MS} &
     (1) & \LImENSB (\LEmENSB) & \LImNTSB (\LEmNTSB) & \LImTESB (\LEmTESB) & \LImETSB (\LEmETSB) \\
   & (2) & \MImENSB (\MEmENSB) & \MImNTSB (\MEmNTSB) & \MImTESB (\MEmTESB) & \MImETSB (\MEmETSB) \\
  \hline                              
  \multirow{2}{*}{On the MS} &
     (1) & \LImENMS (\LEmENMS) & \LImNTMS (\LEmNTMS) & \LImTEMS (\LEmTEMS) & \LImETMS (\LEmETMS) \\
   & (2) & \MImENMS (\MEmENMS) & \MImNTMS (\MEmNTMS) & \MImTEMS (\MEmTEMS) & \MImETMS (\MEmETMS) \\
  \hline                              
  \multirow{2}{*}{Below the MS} &
     (1) & \LImENQU (\LEmENQU) & \LImNTQU (\LEmNTQU) & \LImTEQU (\LEmTEQU) & \LImETQU (\LEmETQU) \\
   & (2) & \MImENQU (\MEmENQU) & \MImNTQU (\MEmNTQU) & \MImTEQU (\MEmTEQU) & \MImETQU (\MEmETQU) \\
   \enddata  
   \tablecomments{     
     The units are in ${\rm 10^9 \, K\,km\,s^{-1}\,pc^2}$ and
     $10^9 \, M_\odot$ for (1) \cotwoone line luminosities (the first
     row in each MS bin) and (2) molecular masses ($M_{\rm gas}$; the
     second row), respectively.  The CO--to--$\rm H_2$ conversion
       factors are assumed to be $\alpha_{\rm CO}=3.6$, 7.2,
       and 10.8
       $\, M_\odot \, ({\rm K\,km\,s^{-1}\,pc^2})^{-1}$ for
       galaxies in the bins of $\log{(M_*/M_\odot)} > 10$,
       $9 < \log{(M_*/M_\odot)} \le 10$, and
       $8 < \log{(M_*/M_\odot)} \le 9$, respectively (see
        \S\ref{subsec:gas_mass}).  The
     values in parentheses are after excluding the galaxies which have
     the \cotwoone line identified by the blind search (see
     \S\ref{subsec:subsamples}).     
   }
\label{tbl:linelum}
\end{deluxetable*}

% Tbl.5
\begin{deluxetable*}{rccccc}
  \tablecaption{Summary of molecular-to-stellar mass ratios
    ($\mu_{\rm gas} = M_{\rm gas}/M_*$) and depletion times
    ($t_{\rm depl} = M_{\rm gas}/SFR$)} \tablehead{
    \colhead{} & \colhead{} & \multicolumn4c{$\log{(M_*/M_\odot)}$} \\
    \colhead{} & \colhead{} & \colhead{$8.0-9.0$} & \colhead{$9.0-10.0$} &
    \colhead{$10.0-11.0$} & \colhead{$11.0-12.0$} } \startdata
  \multirow{2}{*}{Above the MS} &
    (1) & \MMImENSB (\MMEmENSB) & \MMImNTSB (\MMEmNTSB) & \MMImTESB (\MMEmTESB) & \MMImETSB (\MMEmETSB) \\
  & (2) & \DImENSB  (\DEmENSB) &  \DImNTSB  (\DEmNTSB) &  \DImTESB  (\DEmTESB) &  \DImETSB  (\DEmETSB) \\
  \hline                              
  \multirow{2}{*}{On the MS} &
    (1) & \MMImENMS (\MMEmENMS) & \MMImNTMS (\MMEmNTMS) & \MMImTEMS (\MMEmTEMS) & \MMImETMS (\MMEmETMS) \\
  & (2) & \DImENMS  (\DEmENMS) &  \DImNTMS  (\DEmNTMS) &  \DImTEMS  (\DEmTEMS) &  \DImETMS  (\DEmETMS) \\
  \hline                              
  \multirow{2}{*}{Below the MS} &
    (1) & \MMImENQU (\MMEmENQU) & \MMImNTQU (\MMEmNTQU) & \MMImTEQU (\MMEmTEQU) & \MMImETQU (\MMEmETQU) \\
  & (2) & \DImENQU  (\DEmENQU) &  \DImNTQU  (\DEmNTQU) &  \DImTEQU  (\DEmTEQU) &  \DImETQU  (\DEmETQU) \\
  \enddata
  \tablecomments{ The first and second rows in each MS bin are (1)
    molecular-to-stellar mass ratios ($M_{\rm gas}/M_*$) and (2)
    depletion times ($M_{\rm gas}/SFR$ in $\rm Gyr$), respectively.
    The CO--to--$\rm H_2$ conversion
       factors are assumed to be $\alpha_{\rm CO}=3.6$, 7.2,
       and 10.8
       $\, M_\odot \, ({\rm K\,km\,s^{-1}\,pc^2})^{-1}$ for
       galaxies in the bins of $\log{(M_*/M_\odot)} > 10$,
       $9 < \log{(M_*/M_\odot)} \le 10$, and
       $8 < \log{(M_*/M_\odot)} \le 9$, respectively (see
        \S\ref{subsec:gas_mass}). The values in parentheses
    are after excluding the galaxies which have the \cotwoone line
    identified by the blind search (see \S\ref{subsec:subsamples}).}
\label{tbl:M_H2-M*_T_depl}
\end{deluxetable*}

\clearpage

\appendix

\section{The individual spectra in the stacking}\label{app:indiv}

\begin{figure}
  \begin{center}
    \includegraphics[angle=0,width=0.49\textwidth]
    {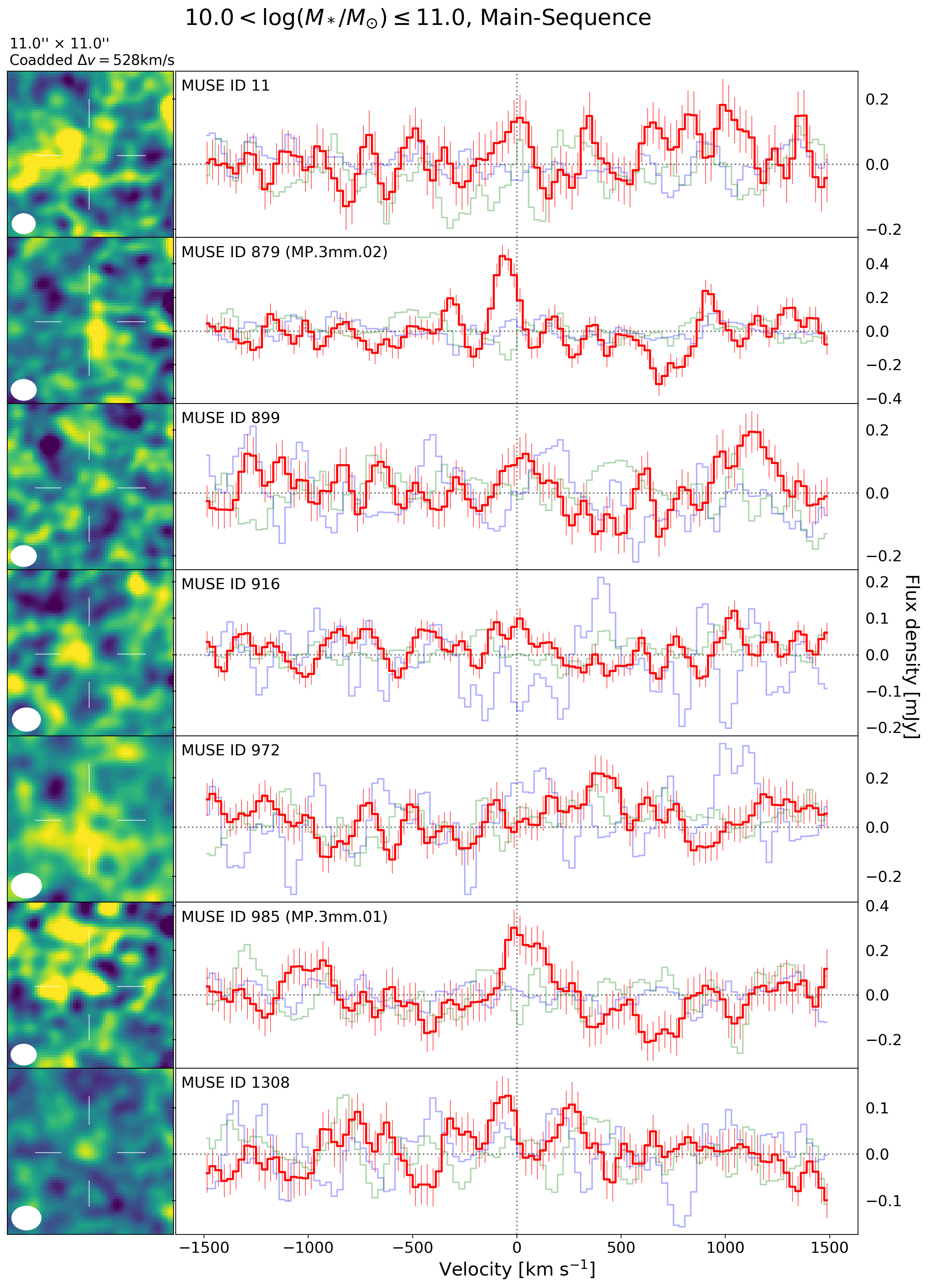}
    \vspace*{5mm}
    \includegraphics[angle=0,width=0.49\textwidth]
    {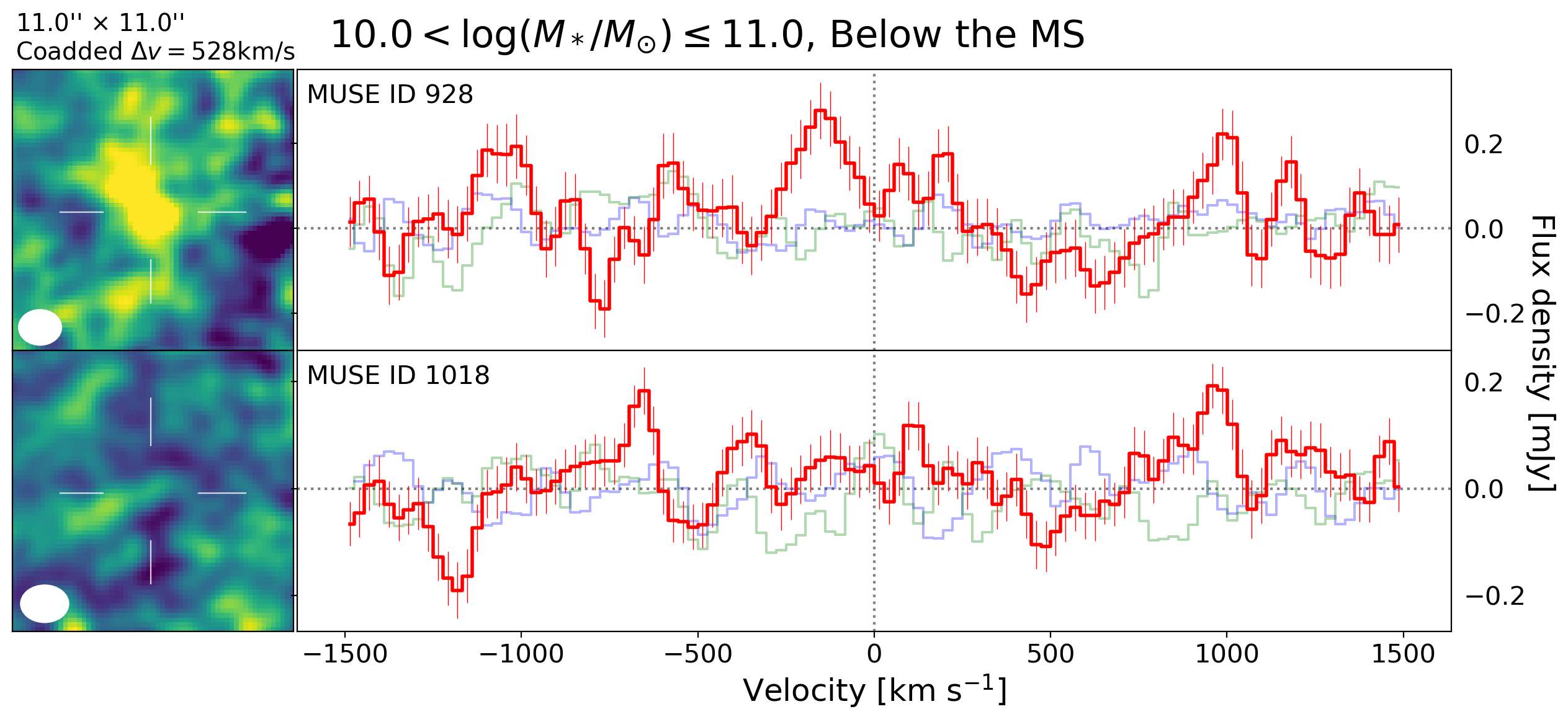}
    \caption{ Individual spectra used for stacking in the bins of
      $10.0 < \log{(M*/M_\odot)} \leq 11.0$ for the galaxies on the MS
      galaxies (left) and below the MS (right). The 2D image
      (moment-0) and 1D spectrum are displayed for each galaxy with
      the MUSE ID number at the top left in each panel of the 1D
      spectrum. If the emission has been detected by the blind
      \citep{Gonz19} or prior searches \citep{Boog19}, then its ASPECS
      ID is also shown in parentheses. The lines and symbols are the
      same as in Figure~\ref{fig:M8-12_2D1Dstack}. }
    \label{fig:indiv_spec}
  \end{center}
\end{figure}

The individual spectra for the stacking of the galaxies on and
  below the MS with $10.0 < \log{(M*/M_\odot)} \leq 11.0$
are displayed in Figure~\ref{fig:indiv_spec} (\S\ref{subsec:indiv}).

% Fig.9

\section{The stacked images, spectra, and measured upper limits for
  high-$J$ CO lines}\label{app:hi_j}

\begin{figure}
  \begin{center}
    \includegraphics[angle=0,trim=0 0 0 40,clip,width=0.48\textwidth]
    {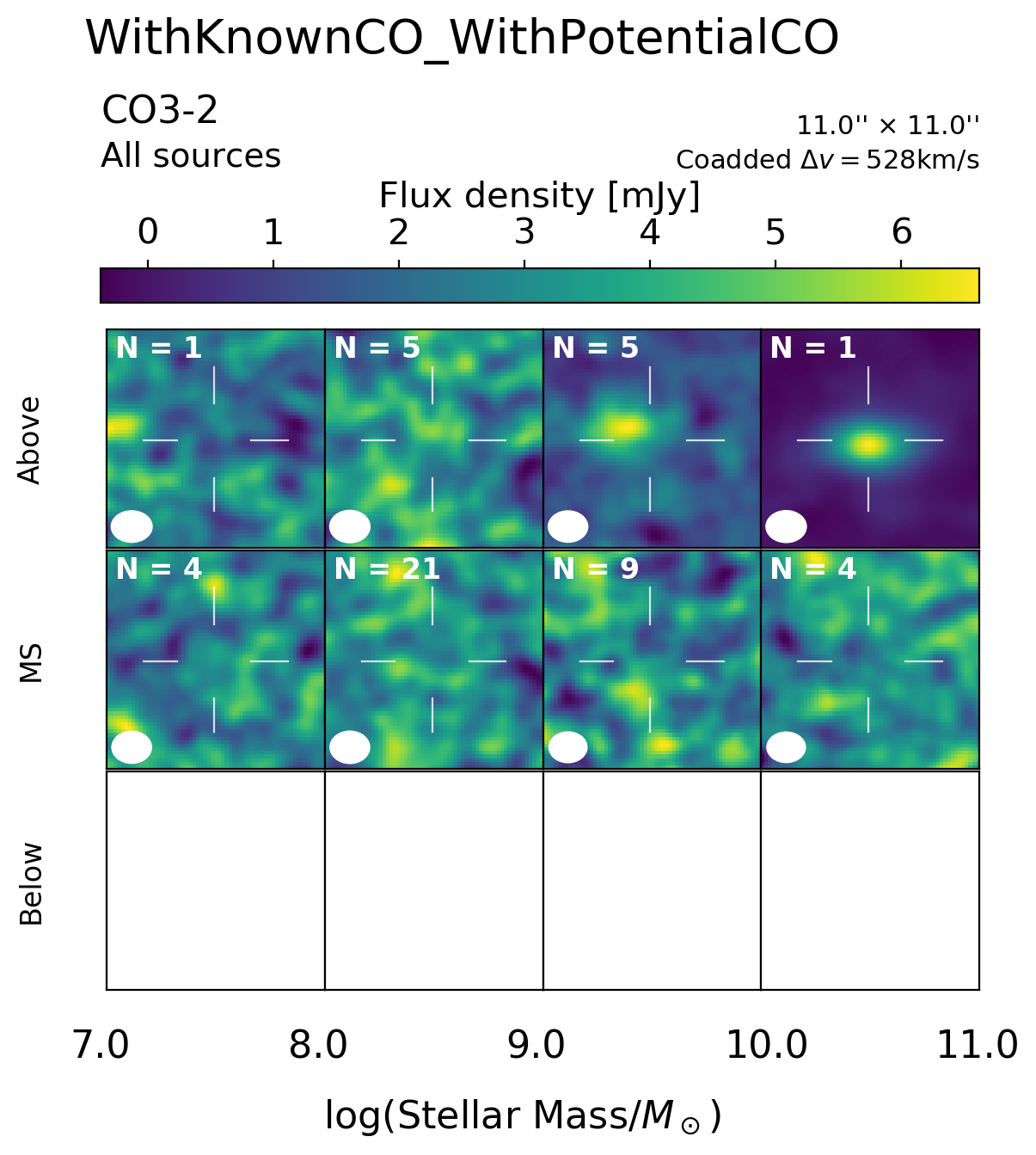}
    \includegraphics[angle=0,trim=0 0 0 40,clip,width=0.48\textwidth]
    {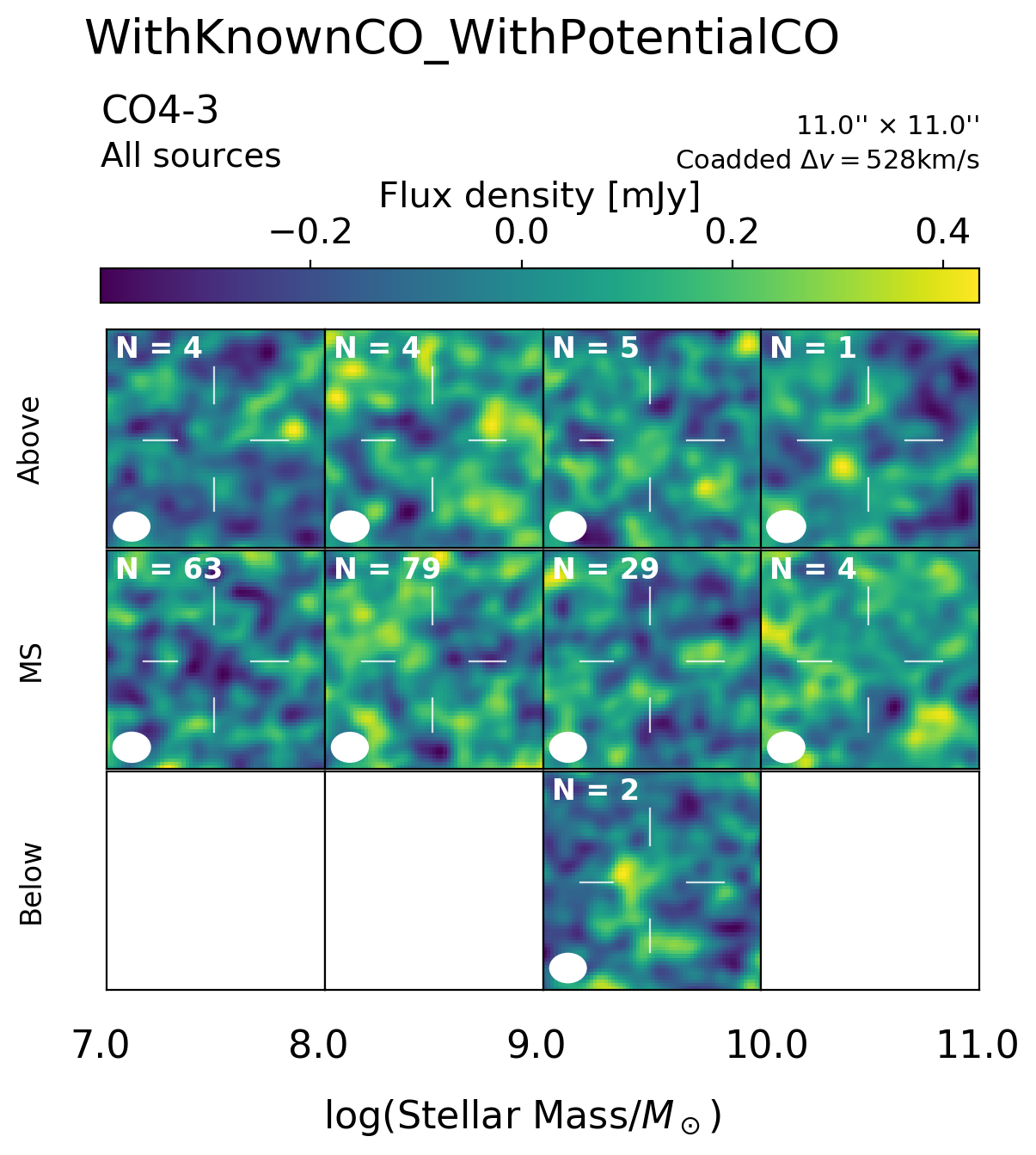}
    \includegraphics[angle=0,trim=0 0 0 30,clip,width=0.48\textwidth]
    {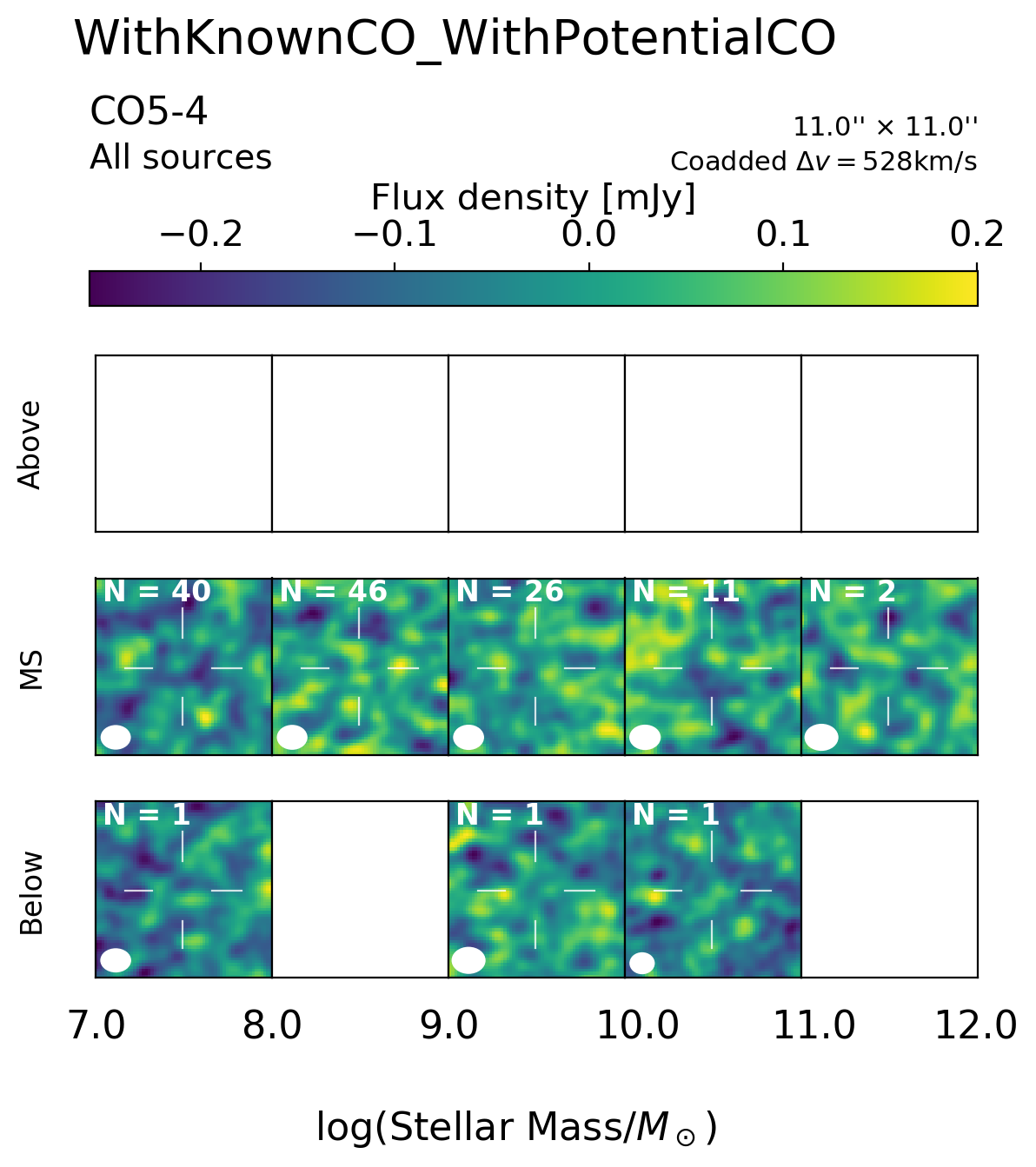}
    \includegraphics[angle=0,trim=0 0 0 30,clip,width=0.48\textwidth]
    {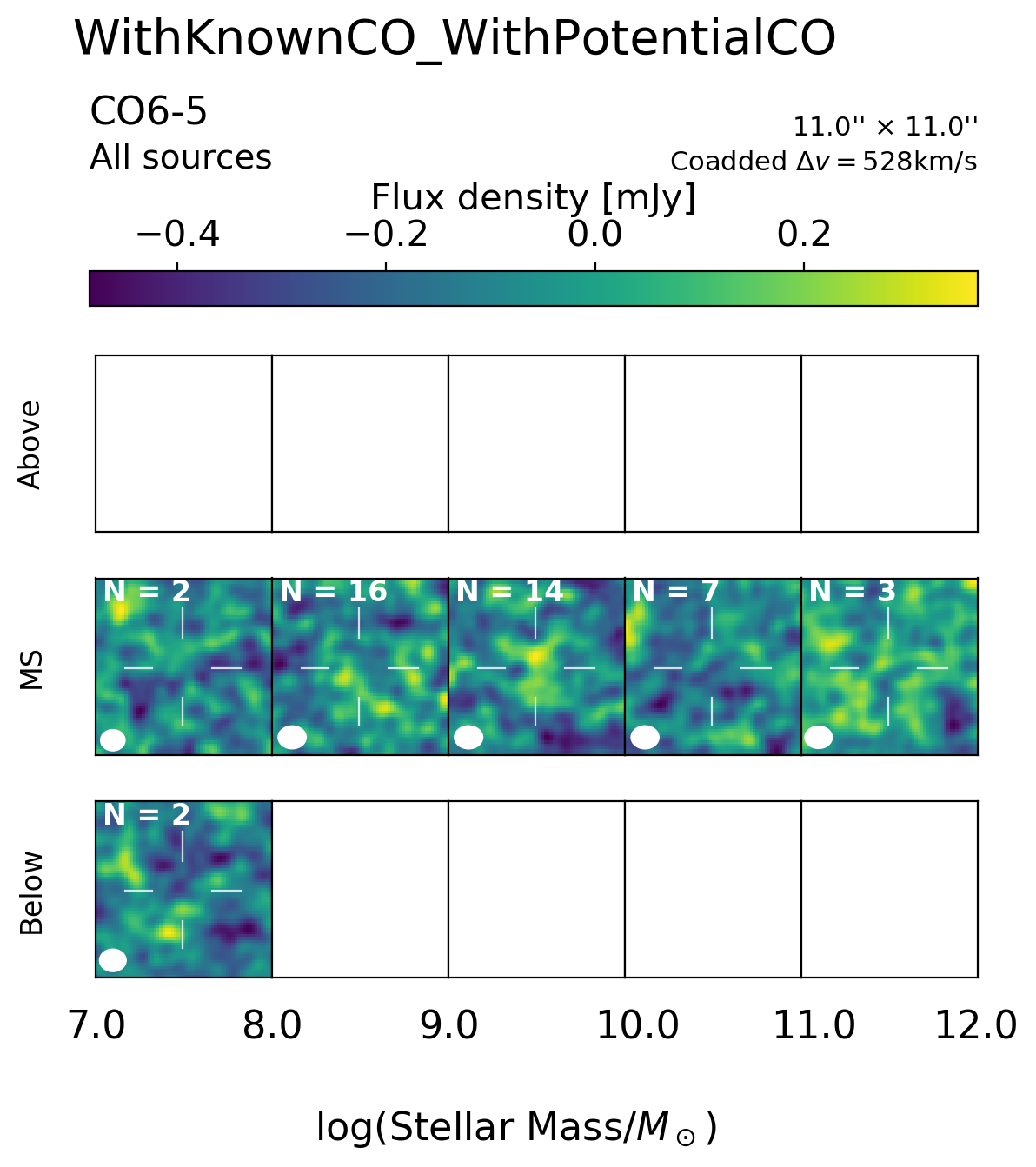}
    \caption{ The stacked results of CO(3-2), CO(4-3), CO(5-4), and
      CO(6-5), shown in the same manner as the left panel in
      Figure~\ref{fig:M8-12_2D1Dstack}, at top left, top right, bottom
      left, and bottom right, respectively. No significant detections
      in the stacks are reported.}
    \label{fig:hi_j}
  \end{center}
\end{figure}

The stacked images and spectra for CO(3-2), CO(4-3), CO(5-4),
and CO(6-5) are shown in Figure~\ref{fig:hi_j}. The measured line
flux upper limits are summarized in Table~\ref{tbl:hi_j_flux}.

% Fig.10

\section{Tentative detections of CO(5-4)}\label{app:CO5-4}

\begin{figure}
  \begin{center}
    \includegraphics[angle=0,trim=0 0 0 0,clip,width=\textwidth]
    {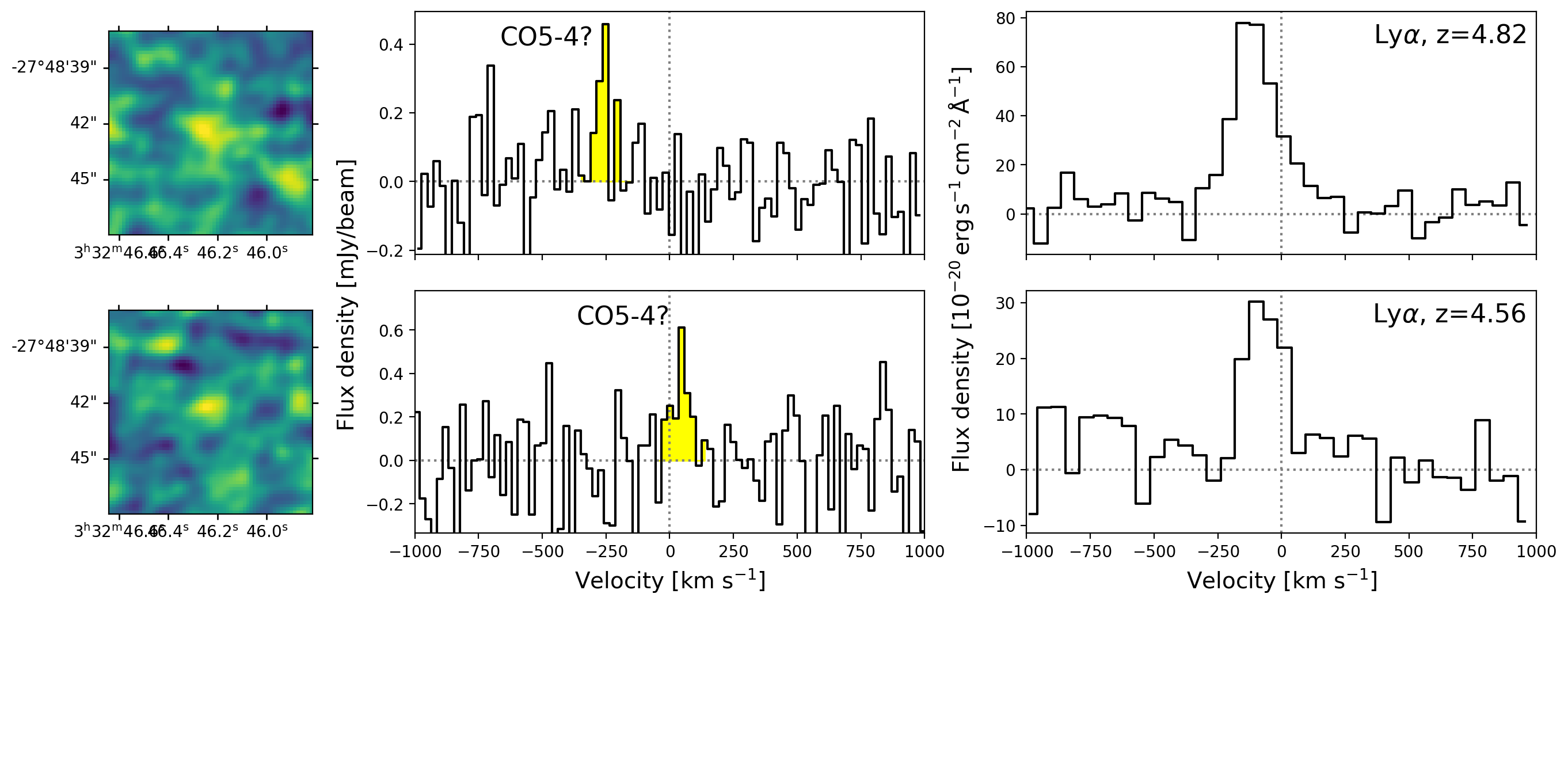}
    \caption{ Tentative detections of CO(5-4) of two Ly$\alpha$
      emitters. The left panels show the 2D images of CO(5-4) created
      over the velocity range highlighted in the middle panels. The
      middle panels depict the ALMA 1D spectra. In the right panels,
      the MUSE 1D spectra are presented along with the MUSE redshift
      in the top right corner. The velocity is defined relative to the
      MUSE redshift with an empirical correction for Ly$\alpha$
      systemic redshift \citep[see
      \S\ref{subsec:results_hiJ};][]{Verh18}.}
    \label {fig:CO5-4}
  \end{center}
\end{figure}

In Figure~\ref{fig:CO5-4}, we present the two cases of potential
CO(5-4) detections. As discussed in \S~\ref{subsec:results_hiJ},
further observations are required to verify these tentative
detections.

% Fig.11

\clearpage

\section{Comparisons with earlier studies}\label{app:Spea14}

\begin{figure*}
  \begin{center}
    \includegraphics[angle=0,width=\textwidth]
    {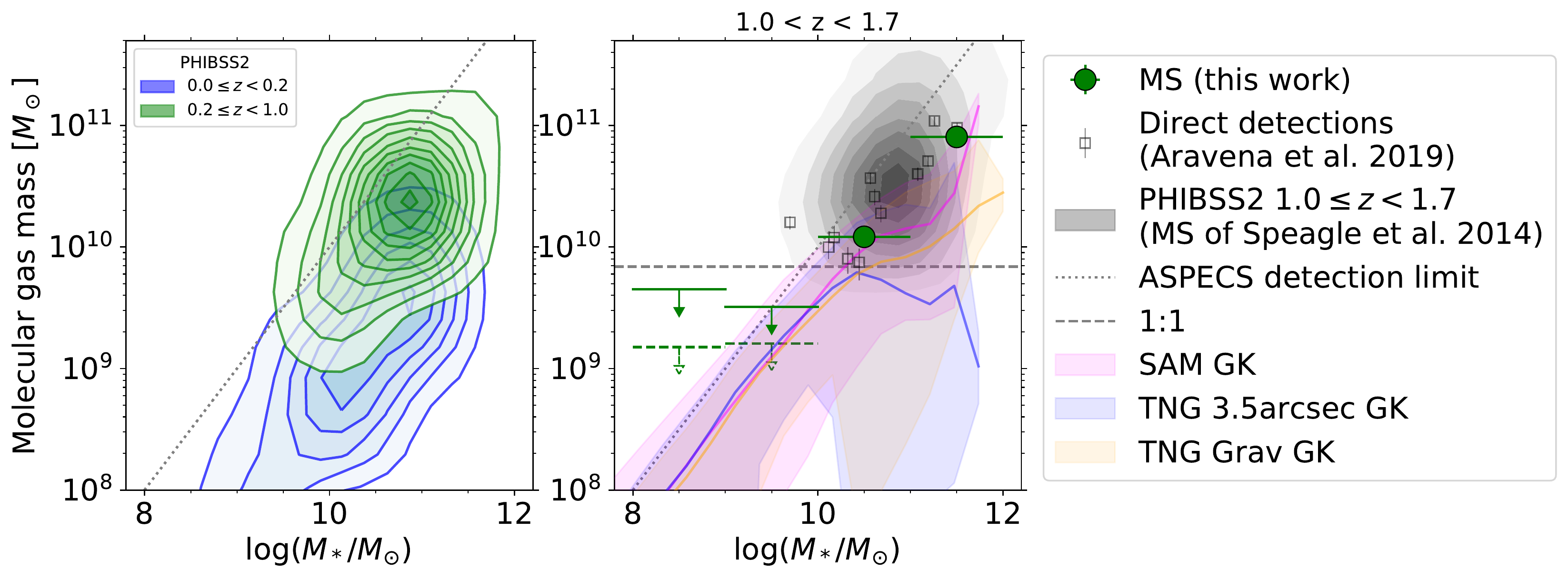}
    \includegraphics[angle=0,width=\textwidth]
    {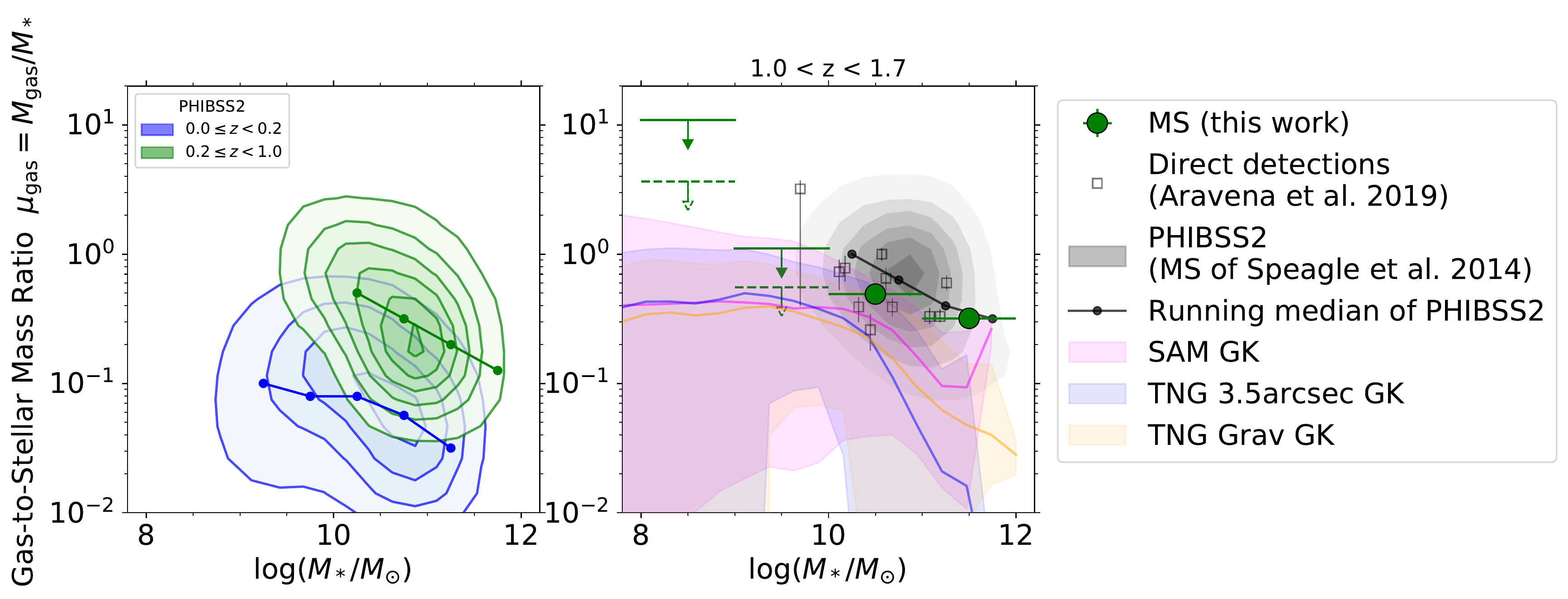}
    \caption{The same diagrams as Figure~\ref{fig:Mgas}, to show our
      results in the context of previous studies of the molecular gas
      content. Note that the MS galaxies here are classified based
        on the prescription of \cite{Spea14} following
        \cite{Tacc18}. Thus, the data points and the MS boundaries are
        slightly shifted compared to Figure~\ref{fig:Mgas} in the
        main text. The distributions of PHIBSS2 galaxies at lower
      redshifts are displayed in the left panels (color-coded by
      redshift as given in the legend). Most of the PHIBSS2 galaxies
      at $z\sim0$ are from xCOLD GASS \citep{Sain17}.  }
    \label{fig:gas_T18_MS}
  \end{center}
\end{figure*}

\begin{figure*}
  \begin{center}
    \includegraphics[angle=0,width=\textwidth]
    {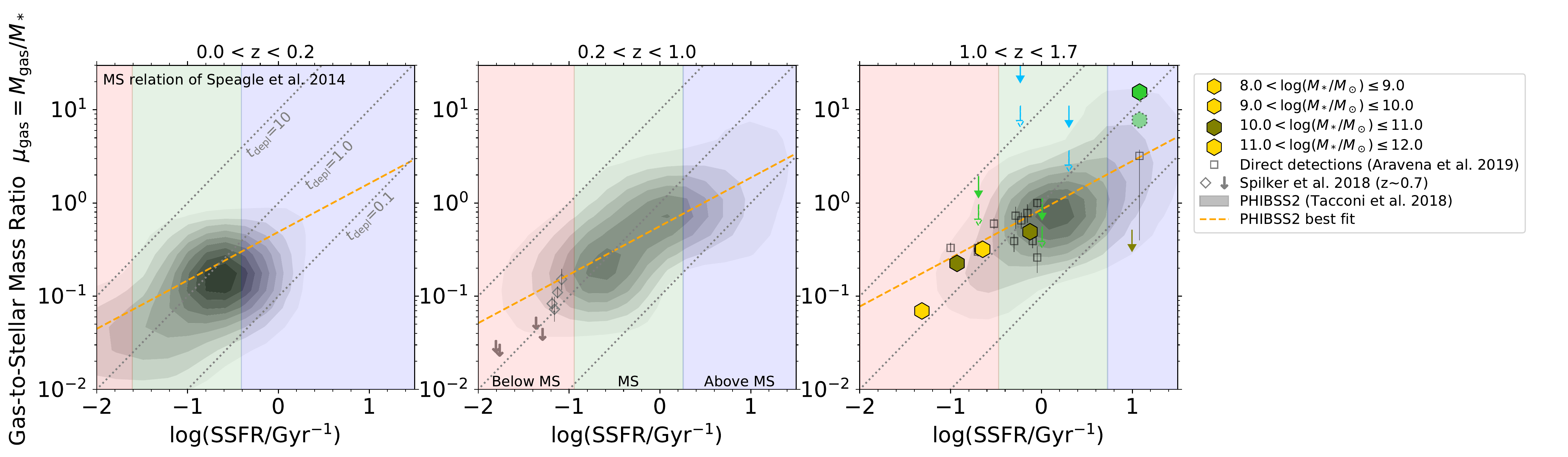}\\
    \includegraphics[angle=0,width=\textwidth]
    {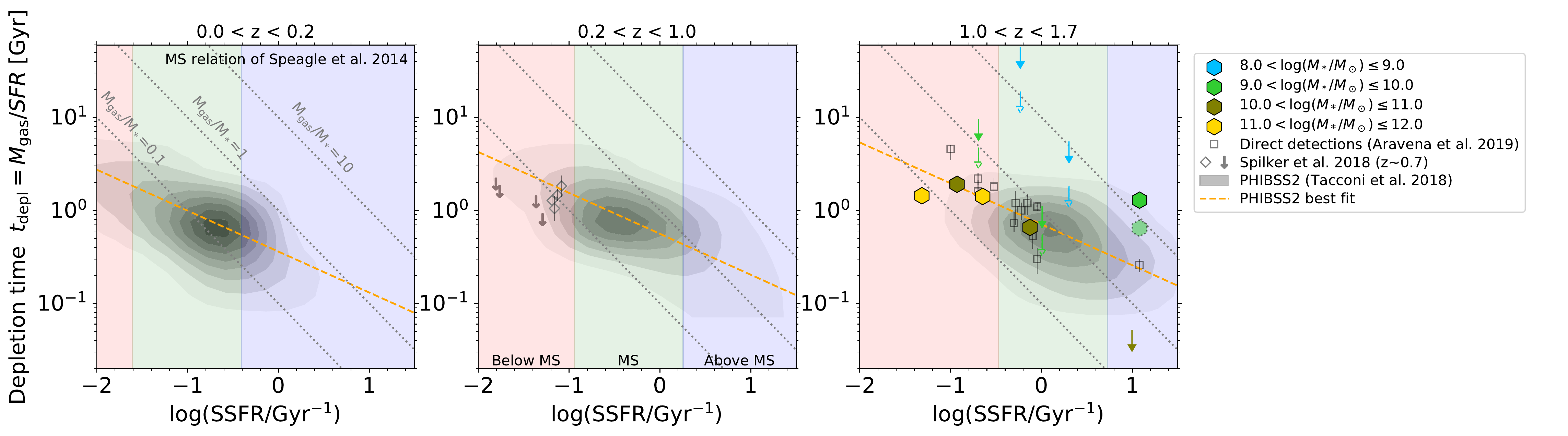}
    \caption{The same diagrams as Figures~\ref{fig:fgas-SSFR} and
      \ref{fig:Tdepl-SSFR} in the main text, but lower redshift bins
      are also shown, following Figure~\ref{fig:gas_T18_MS}. 
        Again, the main-sequence relation of \cite{Spea14} is used
        here instead. Thus, the data points and the MS boundaries are
        slightly shifted compared to Figures~\ref{fig:fgas-SSFR} and
        \ref{fig:Tdepl-SSFR}.  The light pink, green, and blue
      background colors indicate the regions of below, on, and above
      the main-sequence relation at $\log{(M_*/M_\odot)} = 10$,
      respectively. The orange dashed lines are the best-fit lines for
      the PHIBSS2 sample \citep{Tacc18}.  }
    \label{fig:gas_T18_All}
  \end{center}
\end{figure*}

Here we demonstrate our results in the context of earlier studies
including molecular gas measurements for galaxies at lower redshifts.
We adopt the main-sequence relation of \cite{Spea14}, which was also
used in the analysis of PHIBSS2 \citep{Tacc18}, to be consistent with
the literature.  Note that in the main text of this paper, the
main-sequence relation of \cite{Boog18} is utilized throughout,
because this is calibrated based on the MUSE-detected sources whose
average stellar masses and SFRs are lower than the sample used in
\cite{Spea14}. Therefore, the data points and the boundaries of
  the main-sequence relation shown in this Appendix are slightly shifted
  compared to the ones shown in the main text. As presented in the
rightmost panels of Figures~\ref{fig:gas_T18_MS} and
\ref{fig:gas_T18_All}, the results are consistent even when the MS
relation of \cite{Spea14} is adopted.

In this study, we put constraints on the molecular gas content through
CO emission at $\log{(M_*/M_\odot)} \sim 9.5$, which has only been
explored at lower redshift, as illustrated in the two left panels of
Figure~\ref{fig:gas_T18_MS}. The majority of the PHIBSS2 sample at
$z\sim0$ are from xCOLD GASS \citep{Sain11a, Sain11b, Sain16, Sain17}.
We find that at $z\sim1.5$, the molecular gas mass of the MS galaxies
is increasing with increasing stellar mass, at least from
$\log{(M_*/M_\odot)} \sim 9.5$ (top right). At lower redshift, this
trend is seen starting from $\log{(M_*/M_\odot)} \sim 9.0$, but at
smaller gas masses \citep[top left panel; see also][]{Sain17}. In the
bottom panels, a constant gas-to-stellar mass ratio between
$\log{(M_*/M_\odot)} \sim 9.5$ and $\sim 10.5$ is shown at both
$z\sim0$ and $z\sim1.5$ for galaxies lying on the MS. Then the
gas-to-stellar mass ratio declines towards higher stellar masses.

In Figure~\ref{fig:gas_T18_All}, we show the same plots as in
Figures~\ref{fig:fgas-SSFR} and \ref{fig:Tdepl-SSFR} in the main text,
but now the main-sequence relation of \cite{Spea14} is adopted.
We also include the best-fit lines from \cite{Tacc18}.

% Fig.12

% Fig.13

\begin{deluxetable*}{rcccccc}
\tablecaption{Estimated line flux upper limits of the stacked high-$J$ CO emission}
\tablehead{
  \colhead{} & \multicolumn5c{$\log{(M_*/M_\odot)}$} \\
  \colhead{} & \colhead{CO} &
   \colhead{$7.0-8.0$} &   \colhead{$8.0-9.0$} &
  \colhead{$9.0-10.0$} & \colhead{$10.0-11.0$} & \colhead{$11.0-12.0$} }
\startdata
\multirow{4}{*}{Above the MS} & 3-2 & \FTTEmSESB & \FTTEmENSB & \FTTEmNTSB & \FTTEmTESB & \FTTEmETSB \\
                              & 4-3 & \FFTEmSESB & \FFTEmENSB & \FFTEmNTSB & \FFTEmTESB & \FFTEmETSB \\
                              & 5-4 & \FFFEmSESB & \FFFEmENSB & \FFFEmNTSB & \FFFEmTESB & \FFFEmETSB \\
                              & 6-5 & \FSFEmSESB & \FSFEmENSB & \FSFEmNTSB & \FSFEmTESB & \FSFEmETSB \\
\hline                              
   \multirow{4}{*}{On the MS} & 3-2 & \FTTEmSEMS & \FTTEmENMS & \FTTEmNTMS & \FTTEmTEMS & \FTTEmETMS \\
                              & 4-3 & \FFTEmSEMS & \FFTEmENMS & \FFTEmNTMS & \FFTEmTEMS & \FFTEmETMS \\
                              & 5-4 & \FFFEmSEMS & \FFFEmENMS & \FFFEmNTMS & \FFFEmTEMS & \FFFEmETMS \\
                              & 6-5 & \FSFEmSEMS & \FSFEmENMS & \FSFEmNTMS & \FSFEmTEMS & \FSFEmETMS \\
\hline                              
\multirow{4}{*}{Below the MS} & 3-2 & \FTTEmSEQU & \FTTEmENQU & \FTTEmNTQU & \FTTEmTEQU & \FTTEmETQU \\
                              & 4-3 & \FFTEmSEQU & \FFTEmENQU & \FFTEmNTQU & \FFTEmTEQU & \FFTEmETQU \\
                              & 5-4 & \FFFEmSEQU & \FFFEmENQU & \FFFEmNTQU & \FFFEmTEQU & \FFFEmETQU \\
                              & 6-5 & \FSFEmSEQU & \FSFEmENQU & \FSFEmNTQU & \FSFEmTEQU & \FSFEmETQU \\
\enddata
\tablecomments{The units are in ${\rm Jy\,km\,s^{-1}}$.}
\label{tbl:hi_j_flux}
\end{deluxetable*}

\clearpage

\bibliography{bib}

% \begin{thebibliography}{}

% \bibitem[Astropy Collaboration et al.(2013)]{2013A&A...558A..33A} Astropy Collaboration, Robitaille, T.~P., Tollerud, E.~J., et al.\ 2013, \aap, 558, A33 
% \bibitem[Bertin \& Arnouts(1996)]{1996A&AS..117..393B} Bertin, E., \& Arnouts, S.\ 1996, \aaps, 117, 393 
% \bibitem[Corrales(2015)]{2015ApJ...805...23C} Corrales, L.\ 2015, \apj, 805, 23
% \bibitem[Ferland et al.(2013)]{2013RMxAA..49..137F} Ferland, G.~J., Porter, R.~L., van Hoof, P.~A.~M., et al.\ 2013, \rmxaa, 49, 137
% \bibitem[Hanisch \& Biemesderfer(1989)]{1989BAAS...21..780H} Hanisch, R.~J., \& Biemesderfer, C.~D.\ 1989, \baas, 21, 780 
% \bibitem[Lamport(1994)]{lamport94} Lamport, L. 1994, LaTeX: A Document Preparation System, 2nd Edition (Boston, Addison-Wesley Professional)
% \bibitem[Schwarz et al.(2011)]{2011ApJS..197...31S} Schwarz, G.~J., Ness, J.-U., Osborne, J.~P., et al.\ 2011, \apjs, 197, 31  
% \bibitem[Vogt et al.(2014)]{2014ApJ...793..127V} Vogt, F.~P.~A., Dopita, M.~A., Kewley, L.~J., et al.\ 2014, \apj, 793, 127  

% \end{thebibliography}



\end{document}

%% file: CO2-1_table_values.tex
\newcommand{\FImENQU}{--\xspace}     %QU
\newcommand{\FImENMS}{$<0.022$\xspace}     %MS
\newcommand{\FImENSB}{$<0.032$\xspace}     %SB
\newcommand{\FImNTQU}{$<0.051$\xspace}     %QU
\newcommand{\FImNTMS}{$<0.019$\xspace}     %MS
\newcommand{\FImNTSB}{$0.110 \pm 0.027$\xspace}     %SB
\newcommand{\FImTEQU}{$0.119 \pm 0.039$\xspace}     %QU
\newcommand{\FImTEMS}{$0.130 \pm 0.021$\xspace}     %MS
\newcommand{\FImTESB}{$<0.081$\xspace}     %SB
\newcommand{\FImETQU}{$0.152 \pm 0.027$\xspace}     %QU
\newcommand{\FImETMS}{$0.453 \pm 0.042$\xspace}     %MS
\newcommand{\FImETSB}{--\xspace}     %SB
\newcommand{\FEmENQU}{--\xspace}     %QU
\newcommand{\FEmENMS}{$<0.022$\xspace}     %MS
\newcommand{\FEmENSB}{$<0.032$\xspace}     %SB
\newcommand{\FEmNTQU}{$<0.051$\xspace}     %QU
\newcommand{\FEmNTMS}{$<0.019$\xspace}     %MS
\newcommand{\FEmNTSB}{$<0.041$\xspace}     %SB
\newcommand{\FEmTEQU}{$0.119 \pm 0.039$\xspace}     %QU
\newcommand{\FEmTEMS}{$0.095 \pm 0.024$\xspace}     %MS
\newcommand{\FEmTESB}{$<0.081$\xspace}     %SB
\newcommand{\FEmETQU}{$<0.054$\xspace}     %QU
\newcommand{\FEmETMS}{--\xspace}     %MS
\newcommand{\FEmETSB}{--\xspace}     %SB

\newcommand{\LImENQU}{--\xspace}     %QU
\newcommand{\LImENMS}{$<0.43$\xspace}     %MS
\newcommand{\LImENSB}{$<0.62$\xspace}     %SB
\newcommand{\LImNTQU}{$<1.03$\xspace}     %QU
\newcommand{\LImNTMS}{$<0.38$\xspace}     %MS
\newcommand{\LImNTSB}{$2.47 \pm 0.61$\xspace}     %SB
\newcommand{\LImTEQU}{$2.12 \pm 0.69$\xspace}     %QU
\newcommand{\LImTEMS}{$2.51 \pm 0.40$\xspace}     %MS
\newcommand{\LImTESB}{$<1.59$\xspace}     %SB
\newcommand{\LImETQU}{$3.10 \pm 0.55$\xspace}     %QU
\newcommand{\LImETMS}{$10.57 \pm 0.97$\xspace}     %MS
\newcommand{\LImETSB}{--\xspace}     %SB
\newcommand{\LEmENQU}{--\xspace}     %QU
\newcommand{\LEmENMS}{$<0.43$\xspace}     %MS
\newcommand{\LEmENSB}{$<0.62$\xspace}     %SB
\newcommand{\LEmNTQU}{$<1.03$\xspace}     %QU
\newcommand{\LEmNTMS}{$<0.38$\xspace}     %MS
\newcommand{\LEmNTSB}{$<1.02$\xspace}     %SB
\newcommand{\LEmTEQU}{$2.12 \pm 0.69$\xspace}     %QU
\newcommand{\LEmTEMS}{$1.85 \pm 0.47$\xspace}     %MS
\newcommand{\LEmTESB}{$<1.59$\xspace}     %SB
\newcommand{\LEmETQU}{$<1.04$\xspace}     %QU
\newcommand{\LEmETMS}{--\xspace}     %MS
\newcommand{\LEmETSB}{--\xspace}     %SB

\newcommand{\MImENQU}{--\xspace}     %QU
\newcommand{\MImENMS}{$<6.11$\xspace}     %MS
\newcommand{\MImENSB}{$<8.78$\xspace}     %SB
\newcommand{\MImNTQU}{$<9.71$\xspace}     %QU
\newcommand{\MImNTMS}{$<3.62$\xspace}     %MS
\newcommand{\MImNTSB}{$23.42 \pm 5.89$\xspace}     %SB
\newcommand{\MImTEQU}{$10.03 \pm 3.31$\xspace}     %QU
\newcommand{\MImTEMS}{$11.87 \pm 1.95$\xspace}     %MS
\newcommand{\MImTESB}{$<7.53$\xspace}     %SB
\newcommand{\MImETQU}{$14.67 \pm 2.68$\xspace}     %QU
\newcommand{\MImETMS}{$50.07 \pm 5.00$\xspace}     %MS
\newcommand{\MImETSB}{--\xspace}     %SB
\newcommand{\MEmENQU}{--\xspace}     %QU
\newcommand{\MEmENMS}{$<6.11$\xspace}     %MS
\newcommand{\MEmENSB}{$<8.78$\xspace}     %SB
\newcommand{\MEmNTQU}{$<9.71$\xspace}     %QU
\newcommand{\MEmNTMS}{$<3.62$\xspace}     %MS
\newcommand{\MEmNTSB}{$<9.70$\xspace}     %SB
\newcommand{\MEmTEQU}{$10.03 \pm 3.31$\xspace}     %QU
\newcommand{\MEmTEMS}{$8.74 \pm 2.26$\xspace}     %MS
\newcommand{\MEmTESB}{$<7.53$\xspace}     %SB
\newcommand{\MEmETQU}{$<4.95$\xspace}     %QU
\newcommand{\MEmETMS}{--\xspace}     %MS
\newcommand{\MEmETSB}{--\xspace}     %SB

\newcommand{\MMImENQU}{--\xspace}     %QU
\newcommand{\MMImENMS}{$<14.12$\xspace}     %MS
\newcommand{\MMImENSB}{$<22.01$\xspace}     %SB
\newcommand{\MMImNTQU}{$<2.82$\xspace}     %QU
\newcommand{\MMImNTMS}{$<1.25$\xspace}     %MS
\newcommand{\MMImNTSB}{$9.94 \pm 2.50$\xspace}     %SB
\newcommand{\MMImTEQU}{$0.18 \pm 0.06$\xspace}     %QU
\newcommand{\MMImTEMS}{$0.49 \pm 0.08$\xspace}     %MS
\newcommand{\MMImTESB}{$<0.51$\xspace}     %SB
\newcommand{\MMImETQU}{$0.08 \pm 0.01$\xspace}     %QU
\newcommand{\MMImETMS}{$0.24 \pm 0.02$\xspace}     %MS
\newcommand{\MMImETSB}{--\xspace}     %SB
\newcommand{\MMEmENQU}{--\xspace}     %QU
\newcommand{\MMEmENMS}{$<14.12$\xspace}     %MS
\newcommand{\MMEmENSB}{$<22.01$\xspace}     %SB
\newcommand{\MMEmNTQU}{$<2.82$\xspace}     %QU
\newcommand{\MMEmNTMS}{$<1.25$\xspace}     %MS
\newcommand{\MMEmNTSB}{$<4.52$\xspace}     %SB
\newcommand{\MMEmTEQU}{$0.18 \pm 0.06$\xspace}     %QU
\newcommand{\MMEmTEMS}{$0.46 \pm 0.12$\xspace}     %MS
\newcommand{\MMEmTESB}{$<0.51$\xspace}     %SB
\newcommand{\MMEmETQU}{$<0.02$\xspace}     %QU
\newcommand{\MMEmETMS}{--\xspace}     %MS
\newcommand{\MMEmETSB}{--\xspace}     %SB

\newcommand{\DImENQU}{--\xspace}     %QU
\newcommand{\DImENMS}{$<11.27$\xspace}     %MS
\newcommand{\DImENSB}{$<5.95$\xspace}     %SB
\newcommand{\DImNTQU}{$<31.44$\xspace}     %QU
\newcommand{\DImNTMS}{$<1.46$\xspace}     %MS
\newcommand{\DImNTSB}{$1.82 \pm 0.46$\xspace}     %SB
\newcommand{\DImTEQU}{$1.75 \pm 0.58$\xspace}     %QU
\newcommand{\DImTEMS}{$0.69 \pm 0.11$\xspace}     %MS
\newcommand{\DImTESB}{$<0.05$\xspace}     %SB
\newcommand{\DImETQU}{$3.13 \pm 0.57$\xspace}     %QU
\newcommand{\DImETMS}{$1.13 \pm 0.11$\xspace}     %MS
\newcommand{\DImETSB}{--\xspace}     %SB
\newcommand{\DEmENQU}{--\xspace}     %QU
\newcommand{\DEmENMS}{$<11.27$\xspace}     %MS
\newcommand{\DEmENSB}{$<5.95$\xspace}     %SB
\newcommand{\DEmNTQU}{$<31.44$\xspace}     %QU
\newcommand{\DEmNTMS}{$<1.46$\xspace}     %MS
\newcommand{\DEmNTSB}{$<1.84$\xspace}     %SB
\newcommand{\DEmTEQU}{$1.75 \pm 0.58$\xspace}     %QU
\newcommand{\DEmTEMS}{$0.67 \pm 0.17$\xspace}     %MS
\newcommand{\DEmTESB}{$<0.05$\xspace}     %SB
\newcommand{\DEmETQU}{$<3.56$\xspace}     %QU
\newcommand{\DEmETMS}{--\xspace}     %MS
\newcommand{\DEmETSB}{--\xspace}     %SB

%% file: CO3-2_table_values.tex
\newcommand{\FTTEmSEQU}{--\xspace}     %QU
\newcommand{\FTTEmSEMS}{$<0.039$\xspace}     %MS
\newcommand{\FTTEmSESB}{$<0.076$\xspace}     %SB
\newcommand{\FTTEmENQU}{--\xspace}     %QU
\newcommand{\FTTEmENMS}{$<0.014$\xspace}     %MS
\newcommand{\FTTEmENSB}{$<0.026$\xspace}     %SB
\newcommand{\FTTEmNTQU}{--\xspace}     %QU
\newcommand{\FTTEmNTMS}{$<0.025$\xspace}     %MS
\newcommand{\FTTEmNTSB}{$<0.029$\xspace}     %SB
\newcommand{\FTTEmTEQU}{--\xspace}     %QU
\newcommand{\FTTEmTEMS}{$<0.024$\xspace}     %MS
\newcommand{\FTTEmTESB}{--\xspace}     %SB
\newcommand{\FTTEmETQU}{--\xspace}     %QU
\newcommand{\FTTEmETMS}{--\xspace}     %MS
\newcommand{\FTTEmETSB}{--\xspace}     %SB

%% file: CO4-3_table_values.tex
\newcommand{\FFTEmSEQU}{--\xspace}     %QU
\newcommand{\FFTEmSEMS}{$<0.008$\xspace}     %MS
\newcommand{\FFTEmSESB}{$<0.035$\xspace}     %SB
\newcommand{\FFTEmENQU}{--\xspace}     %QU
\newcommand{\FFTEmENMS}{$<0.010$\xspace}     %MS
\newcommand{\FFTEmENSB}{$<0.041$\xspace}     %SB
\newcommand{\FFTEmNTQU}{$<0.065$\xspace}     %QU
\newcommand{\FFTEmNTMS}{$<0.017$\xspace}     %MS
\newcommand{\FFTEmNTSB}{$<0.035$\xspace}     %SB
\newcommand{\FFTEmTEQU}{--\xspace}     %QU
\newcommand{\FFTEmTEMS}{$<0.034$\xspace}     %MS
\newcommand{\FFTEmTESB}{$<0.066$\xspace}     %SB
\newcommand{\FFTEmETQU}{--\xspace}     %QU
\newcommand{\FFTEmETMS}{--\xspace}     %MS
\newcommand{\FFTEmETSB}{--\xspace}     %SB

%% file: CO5-4_table_values.tex
\newcommand{\FFFEmSEQU}{$<0.057$\xspace}     %QU
\newcommand{\FFFEmSEMS}{$<0.013$\xspace}     %MS
\newcommand{\FFFEmSESB}{--\xspace}     %SB
\newcommand{\FFFEmENQU}{--\xspace}     %QU
\newcommand{\FFFEmENMS}{$<0.011$\xspace}     %MS
\newcommand{\FFFEmENSB}{--\xspace}     %SB
\newcommand{\FFFEmNTQU}{$<0.044$\xspace}     %QU
\newcommand{\FFFEmNTMS}{$<0.014$\xspace}     %MS
\newcommand{\FFFEmNTSB}{--\xspace}     %SB
\newcommand{\FFFEmTEQU}{$<0.153$\xspace}     %QU
\newcommand{\FFFEmTEMS}{$<0.026$\xspace}     %MS
\newcommand{\FFFEmTESB}{--\xspace}     %SB
\newcommand{\FFFEmETQU}{--\xspace}     %QU
\newcommand{\FFFEmETMS}{$<0.037$\xspace}     %MS
\newcommand{\FFFEmETSB}{--\xspace}     %SB

%% file: CO6-5_table_values.tex
\newcommand{\FSFEmSEQU}{$<0.052$\xspace}     %QU
\newcommand{\FSFEmSEMS}{$<0.061$\xspace}     %MS
\newcommand{\FSFEmSESB}{--\xspace}     %SB
\newcommand{\FSFEmENQU}{--\xspace}     %QU
\newcommand{\FSFEmENMS}{$<0.018$\xspace}     %MS
\newcommand{\FSFEmENSB}{--\xspace}     %SB
\newcommand{\FSFEmNTQU}{--\xspace}     %QU
\newcommand{\FSFEmNTMS}{$<0.021$\xspace}     %MS
\newcommand{\FSFEmNTSB}{--\xspace}     %SB
\newcommand{\FSFEmTEQU}{--\xspace}     %QU
\newcommand{\FSFEmTEMS}{$<0.027$\xspace}     %MS
\newcommand{\FSFEmTESB}{--\xspace}     %SB
\newcommand{\FSFEmETQU}{--\xspace}     %QU
\newcommand{\FSFEmETMS}{$<0.061$\xspace}     %MS
\newcommand{\FSFEmETSB}{--\xspace}     %SB